\documentclass[twocolumn,superscriptaddress,amsmath,pra,amssymb, aps, nofootinbib]{revtex4-1}
\usepackage{graphicx}
\usepackage{dcolumn}
\usepackage{bm}
\usepackage[colorlinks=true,linkcolor=blue,citecolor=blue,urlcolor=blue]{hyperref}
\usepackage[normalem]{ulem}\usepackage[subrefformat=parens,labelformat=parens,caption=false]{subfig}
\usepackage{verbatim}
\usepackage{cleveref}

\newcommand{\vdp}{vdP}
\newcommand{\ls}{SL}
\newcommand{\dvdp}{DvdP}

\newcommand{\Lcal}{{\cal L}}
\newcommand{\Dcal}{{\cal D}}
\newcommand{\supopdot}{\text{\Large $\cdot$}}

\newcommand{\ann}{\hat{a}}
\newcommand{\adg}{\hat{a}^\dag}
\newcommand{\Hhat}{\hat{H}}

\newcommand{\nn}{\nonumber}

\usepackage{braket}

\begin{document}
\title{Quantum synchronization effects induced by strong nonlinearities}
\author{Yuan Shen}
\affiliation{School of Electrical and Electronic Engineering, Nanyang Technological University, Block S2.1, 50 Nanyang Avenue, Singapore 639798}
\author{Wai-Keong Mok}
\affiliation{Centre for Quantum Technologies, National University of Singapore, Singapore}
\affiliation{California Institute of Technology, Pasadena, CA 91125, USA}
\author{Changsuk Noh}
\affiliation{Department of
Physics, Kyungpook National University, Daegu, South Korea}
\author{Ai Qun Liu}
\email{EAQLiu@ntu.edu.sg}
\affiliation{School of Electrical and Electronic Engineering, Nanyang Technological University, Block S2.1, 50 Nanyang Avenue, Singapore 639798}
\author{Leong-Chuan Kwek}
\email{cqtklc@nus.edu.sg}
\affiliation{Centre for Quantum Technologies, National University of Singapore, Singapore} 
\affiliation{MajuLab, CNRS-UNS-NUS-NTU International Joint Research Unit, Singapore UMI 3654, Singapore}
\affiliation{National Institute of Education, Nanyang Technological University, Singapore 637616, Singapore}
\affiliation{Quantum Science and Engineering Centre (QSec), Nanyang Technological University, Singapore}
\author{Weijun Fan}
\email{EWJFan@ntu.edu.sg}
\affiliation{School of Electrical and Electronic Engineering, Nanyang Technological University, Block S2.1, 50 Nanyang Avenue, Singapore 639798}
\author{Andy Chia}
\affiliation{Centre for Quantum Technologies, National University of Singapore, Singapore}

\begin{abstract}
A paradigm for quantum synchronization is the quantum analog of the Stuart--Landau oscillator, which corresponds to a van~der~Pol oscillator in the limit of weak (i.e.~vanishingly small) nonlinearity. Due to this limitation, the quantum Stuart--Landau oscillator fails to capture interesting nonlinearity-induced phenomena such as relaxation oscillations.
To overcome this deficiency we  propose an alternative model which approximates the Duffing--van~der~Pol oscillator to finitely large nonlinearities while remaining numerically tractable. This allows us to uncover interesting phenomena in the deep-quantum strongly-nonlinear regime with no classical analog, such as the persistence of amplitude death on resonance. We also report nonlinearity-induced position correlations in reactively coupled quantum oscillators. Such coupled oscillations become more and more correlated with increasing nonlinearity before reaching some maximum. Again, this behavior is absent classically. We also show how strong nonlinearity can enlarge the synchronization bandwidth in both single and coupled oscillators. This effect can be harnessed to induce mutual synchronization between two oscillators initially in amplitude death.
\end{abstract}

\maketitle


{\it Introduction.---}Mathematical modelling has shown us how the immense variety and beauty of nature can be governed by nonlinear differential equations~\cite{May01,EK05,JS07,Deb12}. Such equations, owing to their nonlinearity, are difficult to analyze and their application to physical processes has come to be known as nonlinear science~\cite{Cam87,Sco05}.

How nonlinear effects play out in quantum-mechanical systems is a subject of intense investigation. By far, chaos has attracted the most attention~\cite{Sto99,Haa01,Wal12,Wim14}. Nonetheless, noise effects, such as stochastic resonance~\cite{GHJM98,WSB04,Lud19,WTB+19} and coherence resonance~\cite{WSB04,KN21} have also been prominent, while a recent example is Ref.~\cite{SL23}. Another widely recognized effect is noise-induced transitions \cite{HL06}. Recent work has extended this effect to quantum mechanics by using a quantum stochastic differential equation and its correspondence to nonlinear dissipation \cite{CMNK22} (see also Ref.~\cite{CHN+20}). There are also promising applications of nonlinear dissipation such as stabilizing bosonic qubits for fault-tolerant quantum computing~\cite{mirrahimi2014dynamically,leghtas2015confining,chamberland2022building}, and enhancing the sensitivity of quantum sensors~\cite{dutta2019critical,sanchez2021quantum}.

A relative newcomer to the study of nonlinear effects in quantum systems is synchronization~\cite{RP03}. Its most elementary form consists of applying a sinusoidal force, say with amplitude $f$, and frequency $\Omega_{\rm d}$, to a self-sustained oscillator. Synchronization is then the modification of the oscillator frequency to $\Omega_{\rm d}$. A prototypical model is the driven van~der~Pol (vdP) oscillator~\cite{Str15}, defined by phase-space coordinates $(x,y)$ satisfying
\begin{equation}
\label{CvdP}
	x' = y \, ,  \quad
	y' = f \cos(\Omega_{\rm d} t) -\omega_0^2 \, x - \mu \, (x^2 - q^2) \, y  \;,
\end{equation}
where primes denote differentiation with respect to the argument (in this case $t$, representing time). In the absence of forcing ($f=0$) the oscillator is characterized by $\omega_0$, and a nonlinearity parameter $\mu$ which controls how much the oscillator is damped towards an amplitude of order $|q|$. An important feature of the undriven \vdp\ system is the existence of a supercritical Hopf bifurcation at $\mu=0$, via which a stable limit cycle appears for $\mu>0$~\cite{Str15}.

At $\mu=0$, \eqref{CvdP} is entirely linear. This motivates one to consider the quasilinear limit of \eqref{CvdP}, defined by $\mu\longrightarrow0^+$. In this limit the \vdp\ oscillator is well approximated by the Stuart--Landau (SL) oscillator, the steady state of which is rotationally symmetric in phase space (a circular limit cycle). This makes the \ls\ oscillator much simpler to analyze, and has thus served as a starting point in the literature on quantum synchronization for continuous-variable systems, e.g.~Refs.~\cite{lee2013quantum,walter2014quantum,SHM+18,mok2020/physrevresearch.2.033422,lee2014/physreve.89.022913,BOZSB15,WWM17,walter2015quantum,MH15,IK17,AKLB18,BKBB20,BKB21,KYN19,KN20,banerjee2021revival}. The trade-off of course, is that effects taking place at finite values of $\mu$ are excluded. A well known example of this is relaxation oscillations in the undriven \vdp\ oscillator\footnote{In fact, \vdp\ intended for \eqref{CvdP} to model relaxation oscillations in an electrical circuit~\cite{GL12}. To observe relaxation oscillations in quantum theory one needs to quantize the exact \vdp\ model, and it is only relatively recently that such efforts have been made~\cite{SCVC15,chia2020relaxation,ACL21}.}~\cite{JS07,Str15,Jen13}. More effects start to appear if driving is included, such as quasiperiodicity and chaos~\cite{BT11,LR03,Moo04}, both of which are absent in the driven \ls\ oscillator. 

In this work, we investigate the effects of nonlinearity in quantum oscillators by considering a more general model based on the classical Duffing--van~der~Pol (DvdP) oscillator. This adds $\zeta\,x^3$ to $y'$ where $\zeta$ is another nonlinearity parameter. To overcome the inadequacy of the \ls\ model we propose a quantum \dvdp\ oscillator in which the \vdp\ and Duffing nonlinearities (respectively $\mu$ and $\zeta$) are nonvanishing, but also not arbitrarily large. Our model is accurate up to order $(\mu/\omega_0)^2$, at which the distinct signatures of strong nonlinearity appear, such as relaxation oscillations~\cite{chia2020relaxation}. Our approach has the benefit of capturing novel nonlinear effects while evading the large computational cost of simulating quantum systems with very strong nonlinear dissipations. 

We show that for a single oscillator with periodic forcing there exists a critical Duffing nonlinearity, above which further increases in $\zeta$ enlarges the synchronization bandwidth (the amount of detuning the forcing can tolerate from the oscillator and still entrain it). This result is similar to the synchronization enhancement from the classical literature~\cite{antonio2015nonlinearity}, but now generalized to quantum oscillators.\footnote{It is also worth mentioning that nonlinear oscillators are of interest to quantum information too, when they are coupled to qubits. In this context the Duffing nonlinearity has been shown to both increase and stabilize the oscillator-qubit entanglement~\cite{montenegro2014nonlinearity}.} In contrast, the \vdp\ nonlinearity activates genuine quantum effects. Coupling two \vdp\ oscillators dissipatively may lead to either amplitude death (the cessation of oscillations), or mutual synchronization. Classically, amplitude death occurs only when the two oscillators are sufficiently detuned~\cite{SPR12}. Interestingly, we find this need not be the case for quantum oscillators. We show that two quantum \vdp\ oscillators possessing relatively small limit cycles and nonvanishing nonlinearities can exhibit amplitude death even with zero detuning. Larger limit cycles on the other hand can mutually synchronize from a state of amplitude death if their nonlinearity is increased.

We also consider reactively coupled oscillators. Two such \ls\ oscillators cannot develop positional correlations, and hence do not synchronize. This is true regardless of whether the oscillators are classical or quantum. We show here that at finitely large nonlinearity, position correlations behave rather differently between the classical and quantum oscillators: Two reactively coupled quantum \vdp\ oscillators can undergo nonlinearity-induced correlations whereby their position correlation increases as they become more nonlinear. In contrast, we find that making the analogous classical oscillators more nonlinear monotonically reduces their position correlation. The nonlinearity-induced correlations in the quantum \vdp\ oscillators are thus a consequence of both their quantum nature and strong nonlinearity.


{\it Model.---}For simplicity we consider here a dimensionless \dvdp\ model in terms of the nonlinearity parameters $\lambda \equiv \mu q^2 / \omega_0 r^2$ and $\beta \equiv \zeta q^2 / \omega_0^2 r^2$ in which $r$ is a dimensionless scale parameter: 
\begin{equation}
\label{eq:classical dimless duffing-vdp}
    \tilde{x}' = \tilde{y} \,, \quad  
    \tilde{y}'= F \cos(\omega_{\rm d} \tilde{t}\,) -\tilde{x} - \lambda (\tilde{x}-r^2) \tilde{x}' - \beta \, \tilde{x}^3 .
\end{equation}

Note that $\tilde{x} \equiv x r / q$ is now a function of $\tilde{t} = \omega_0 t$, and we have also included a dimensionless external force parameterized by $F = f r / \omega_0^2 q$ and $\omega_{\rm d} = \Omega_{\rm d}/\omega_0$.
From the approximate analysis of \eqref{eq:classical dimless duffing-vdp}, the leading contribution to the oscillator frequency is quadratic in $\lambda$, and linear in $\beta$~(see Appendix), given by $\omega \approx 1 + r^2(3\beta/2 - \lambda^2 r^2/16)$.  This motivates a Bogoliubov--Krylov time-average of the equations of motion up to these orders, giving \cite{sanders2007averaging}
%
\begin{align}
    \alpha' = {}& i\,\frac{F}{2} \cos(\omega_{\rm d} t) -i\,\alpha - i \frac{3\beta}{2} |\alpha|^2 \alpha  + \frac{\lambda}{2} (r^2 - |\alpha|^2) \alpha  \nonumber\\
                 &+ i \frac{\lambda^2}{8} \left( r^4 - 6 r^2 |\alpha|^2 + \frac{11}{2} |\alpha|^4 \right) \alpha, 
                 \label{eq:2ndorder_avg}
\end{align}
where $\alpha = (\tilde{x} + i\,\tilde{y})/2$.
For $F = 0$, \eqref{eq:2ndorder_avg} predicts a limit-cycle amplitude of $2|\alpha| = 2 r$ with the expected frequency shifts due to $\lambda$ and $\beta$. Additionally, note the first-order averaging in $\lambda$ for $\beta = 0$ yields the \ls\ equation. Our approximate model captures the effects of strong \vdp\ nonlinearity of order $\lambda^2$. We seek a quantum master equation $\rho'=\Lcal\rho$ such that $\braket{\hat{a}}'={\rm Tr}[\hat{a}\,\Lcal\rho]$ (with $[\hat{a},\hat{a}^\dag] = \hat{1}$) agrees with~\eqref{eq:2ndorder_avg} in the mean-field limit~\cite{chia2020relaxation}. It can then be shown that this is satisfied by the Lindbladian~\cite{Lin76,GKS76,BP02}
\begin{equation}
\label{L2ndOrder}
    \Lcal = -i \, [\Hhat,\supopdot] + \lambda r^2 \mathcal{D}[\adg] + \frac{\lambda}{2} \, \mathcal{D}[\ann^2] \, ,
\end{equation}
where 
\begin{equation}
\begin{split} 
\label{H02ndOrder}
    \Hhat = {}& \left(1-\frac{\lambda^2 r^4}{8}\right) \adg \ann + \frac{3\lambda^2 r^2}{8} \, \adg{}^2 \ann^2 
                - \frac{11\lambda^2}{48} \, \adg{}^3 \ann^3  \\
              & + \frac{3\beta}{4} \, \adg{}^2 \ann^2 - \frac{F}{2} \cos(\omega_{\rm d} t) (\ann + \adg)\, .
\end{split}
\end{equation}
We have also defined $\Dcal[\hat{c}]\equiv\hat{c}\,\supopdot\,\hat{c}^\dag-(\hat{c}^\dag\hat{c}\,\supopdot + \supopdot\,\hat{c}^\dag\hat{c})/2$ for any $\hat{c}$, and a dot denotes the position of $\rho$ when acted upon by a superoperator. Detailed derivation can be found in the Appendix. We remark that both the higher-order Kerr terms and the nonlinear two-photon dissipation in our proposed model can be implemented in circuit QED~\cite{hillmann2022designing,leghtas2015confining}.
The tunability of the limit cycle radius $r$ allows us to access different parameter regimes of the quantum oscillator, in particular the quantum ($r\ll 1$), and semiclassical ($r\approx1$) regimes. We have included the second-order contributions in $\lambda$ in our model for its nonlinearity-tuning capability, since the terms linear in $\lambda$ neither affect the limit-cycle amplitude nor phase dynamics. The Duffing nonlinearity translates to a Kerr term in $\Lcal$. This model can be considered as an alternative to the  quantum \ls\ oscillator, but with flexibility in tuning the nonlinearity. All our numerical results for a given parameter set are obtained with a sufficiently large truncation of the Hilbert space by ensuring the corresponding steady-state power spectrum converge.

{\it Nonlinearity-enhanced synchronization.---}We study first the frequency locking of the approximate quantum \dvdp\ oscillator to a periodic force [\eqref{L2ndOrder} and \eqref{H02ndOrder}]. The synchronization bandwidth is the range of $\omega_{\rm d}$ for which the oscillator frequency is locked to the driving frequency at steady state. This is achieved when $|\omega_{\rm d} - \tilde{\omega}| = 0$, where $\tilde{\omega}$ the observed frequency of the driven oscillator, obtained from the peak of its spectrum averaged over one period of the drive \cite{chia2020relaxation}.

\begin{figure}[t]
    \centering
    \includegraphics[width=\linewidth]{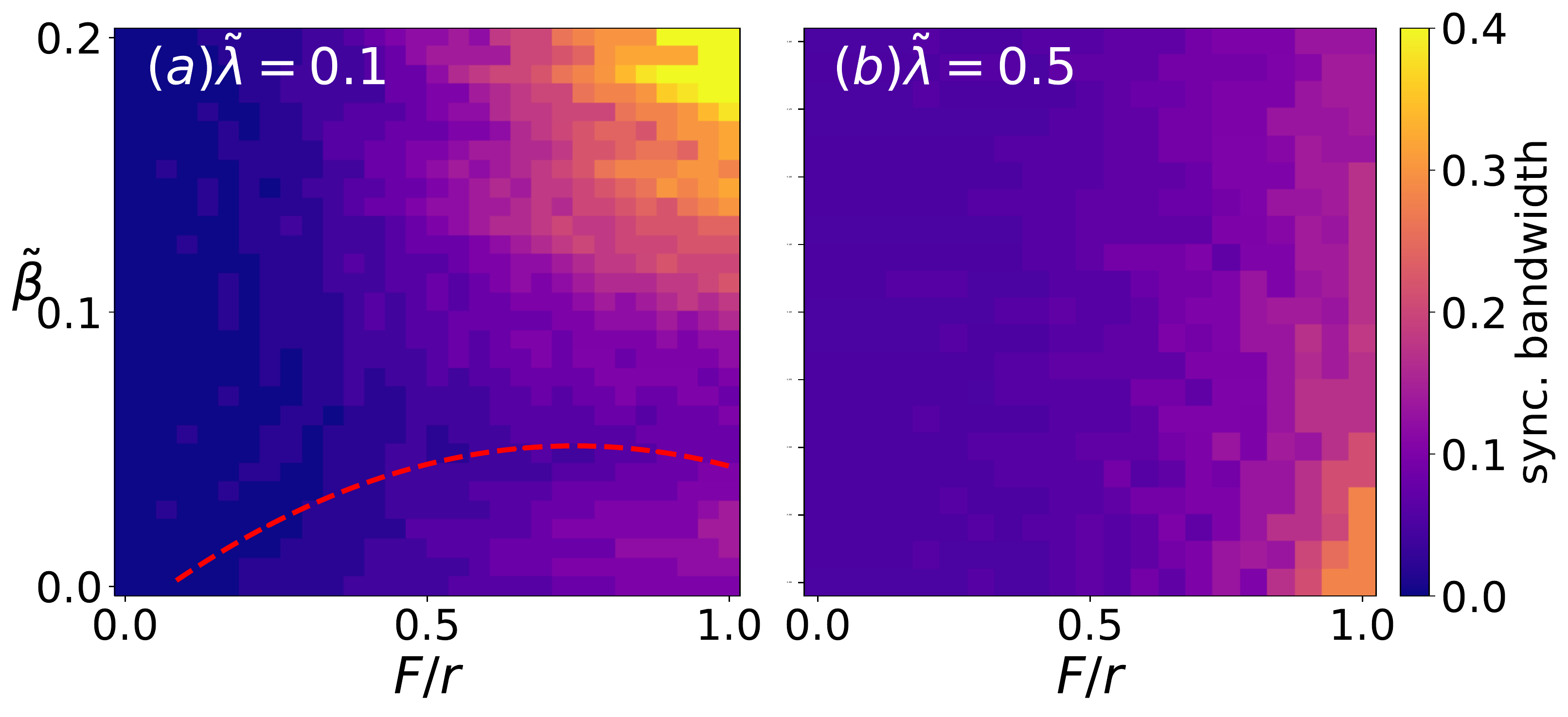}  \caption{Contour plot of the synchronization bandwidth for the quantum \dvdp\ oscillator as a function of $\bar{\beta}\equiv\beta r^2$ (vertical axis) and $F/r$ (horizontal axis) with unit limit-cycle radius, i.e.~$r=1$. In this case $\bar{\beta}=\beta$ and $\bar{\lambda}=\lambda$. The axes are also indicated in subplot (b). (a) Illustration of synchronization enhancement for $\bar{\lambda}=0.1$. Above a critical value of $\bar{\beta}$, indicated by a red dashed line (obtained numerically), the synchronization bandwidth is enlarged as the Duffing nonlinearity is increased. (b) Synchronization enhancement disappears if we increase the \vdp\ nonlinearity from $\bar{\lambda}=0.1$ to $\bar{\lambda}=0.5$, demonstrating the finite range of $\bar{\lambda}$ over which the enhancement is effective.}
    \label{fig:sync bandwidth}
\end{figure}

Here we find the Duffing nonlinearity to enhance quantum synchronization: For a range of $\lambda$ and a fixed $r$, increasing $\beta$ past a critical value widens the synchronization bandwidth linearly. This is illustrated in Fig.~\ref{fig:sync bandwidth}(a) where the synchronization bandwidth is plotted as a contour against $\bar{\beta} \equiv \beta r^2$ and $F/r$. The critical value of $\bar{\beta}$ is indicated by the red dashed line, where the bandwidth is equal to its corresponding value at $\bar{\beta} = 0$. However, this enhancement does not occur for all values of the \vdp\ nonlinearity. In Fig.~\ref{fig:sync bandwidth}(b) we see that an increase of $\bar{\lambda}\equiv\lambda r^2$ from its value in Fig.~\ref{fig:sync bandwidth}(a) can ruin the gain in synchronization bandwidth due to $\bar{\beta}$. Noting that the dissipative terms in $\Lcal$ are all proportional to $\lambda$, this effect can be qualitatively attributed to the phase diffusion due to quantum noise, which is known to inhibit synchronization~\cite{walter2014quantum,lee2014/physreve.89.022913,lee2013quantum}. We can develop some understanding of the quantum \dvdp\ by examining its classical analog. Using the method of harmonic balance on $x$, we are able to derive the conditions for nonlinearity-enhanced synchronization analytically for the classical DvdP oscillator~(see derivation in Appendix), given by
\begin{equation}
    \bar{\beta} > \frac{\bar{\lambda}^2}{3(1-\bar{\lambda}^2)}  \;,  \quad   0 < \bar{\lambda} < 1  \;.
\end{equation}
This shows clearly the existence of a critical value of $\bar{\beta}$, and a finite interval of $\bar{\lambda}$ over which the synchronization enhancement occurs. These results are consistent with Fig.~\ref{fig:sync bandwidth} except for the fact that quantum noise makes the range of $\lambda$ for synchronization enhancement in the quantum \dvdp\ oscillator smaller compared to the classical range as seen in Fig.~\ref{fig:sync bandwidth}(b). 

We have also shown in Appendix that increasing the \vdp\ nonlinearity in the quantum model can only reduce the synchronization bandwidth. One can thus be certain that the synchronization enhancement seen in our model is induced by the Duffing nonlinearity. We shall see next that the \vdp\ nonlinearity can induce synchronization in coupled oscillators from a state of amplitude death.


{\it Nonlinearity-induced synchronization and amplitude death.---}Two dissipatively coupled \vdp\ oscillators (i.e.~no Duffing nonlinearity) can be described by the Lindbladian 

\begin{align}
\label{eq:coupled duffing_vdp me}
    \Lcal = \Lcal_1 + \Lcal_2 -i \, \Delta  [\hat{a}_2^\dag \hat{a}_2,\supopdot]+ \eta \, \Dcal[\ann_1 - \ann_2]  \; ,  
\end{align}
where $\Lcal_k$ ($k=1,2$) is the Lindbladian for oscillator $k$,  defined by setting $\ann$ to $\ann_k$ and $\beta=F=0$ in \eqref{L2ndOrder} and \eqref{H02ndOrder}. We have assumed the oscillators to be identical (i.e.~same $\lambda$ and $r$) except for an initial detuning of $\Delta$, and denoted their coupling strength by $\eta$.

In this case, frequency locking occurs when the observed frequencies of the two oscillators become identical at steady state. As before, the observed frequency of oscillator $k$ is given by the location at which its spectrum is peaked, i.e.~where  $S_{k}(\omega)=\int_{-\infty}^{\infty} dt\,\exp(i\omega t)\langle \adg_{k}(t)\ann_{k}(0)\rangle$ is maximized, and where the two-oscillator steady state is to be used. For a fixed $\eta$, we define the synchronization bandwidth to be the range of $\Delta$ for which the two oscillators lock frequencies. It will also be interesting to look at position correlations in the two oscillators at steady state, defined by
\begin{equation}
\label{PearCorr}
    \Sigma = \frac{\braket{\hat{x}_1 \hat{x}_2}- \braket{\hat{x}_1}\braket{\hat{x}_2}}{\sqrt{\big[\braket{\hat{x}_1^2}-\braket{\hat{x}_1}^2\big]\,\big[\braket{\hat{x}_2^2}-\braket{\hat{x}_2}^2\big]}} \; ,
\end{equation}
where $\hat{x}_k = \ann_k + \adg_k$. Note that frequency locking implies a nonzero $\Sigma$, but not vice versa.

In addition to frequency locking, dissipatively coupled oscillators can also cease to oscillate. If the oscillators are classical, then this may happen for a range of $\eta$ provided that $\Delta$ is sufficiently large. And if both oscillators stabilize to the same phase-space point, which may be taken to be the origin without loss of generality, then the effect is termed amplitude death \cite{SPR12,KVK13}. To define amplitude death in quantum oscillators we generalize the notion of P-bifurcations from classical stochastic systems to the steady-state Wigner function of a reduced state (see Ref.~\cite{CMNK22} and other references therein). In this case, amplitude death is said to occur if the single-oscillator Wigner functions peak at the origin in quantum phase space. This approach is consistent with previous studies on amplitude death in coupled quantum oscillators \cite{BKBB20,AKLB18,BKB21,IK17}.

\begin{figure}[hbt!]
\centering
\includegraphics[width=\linewidth]{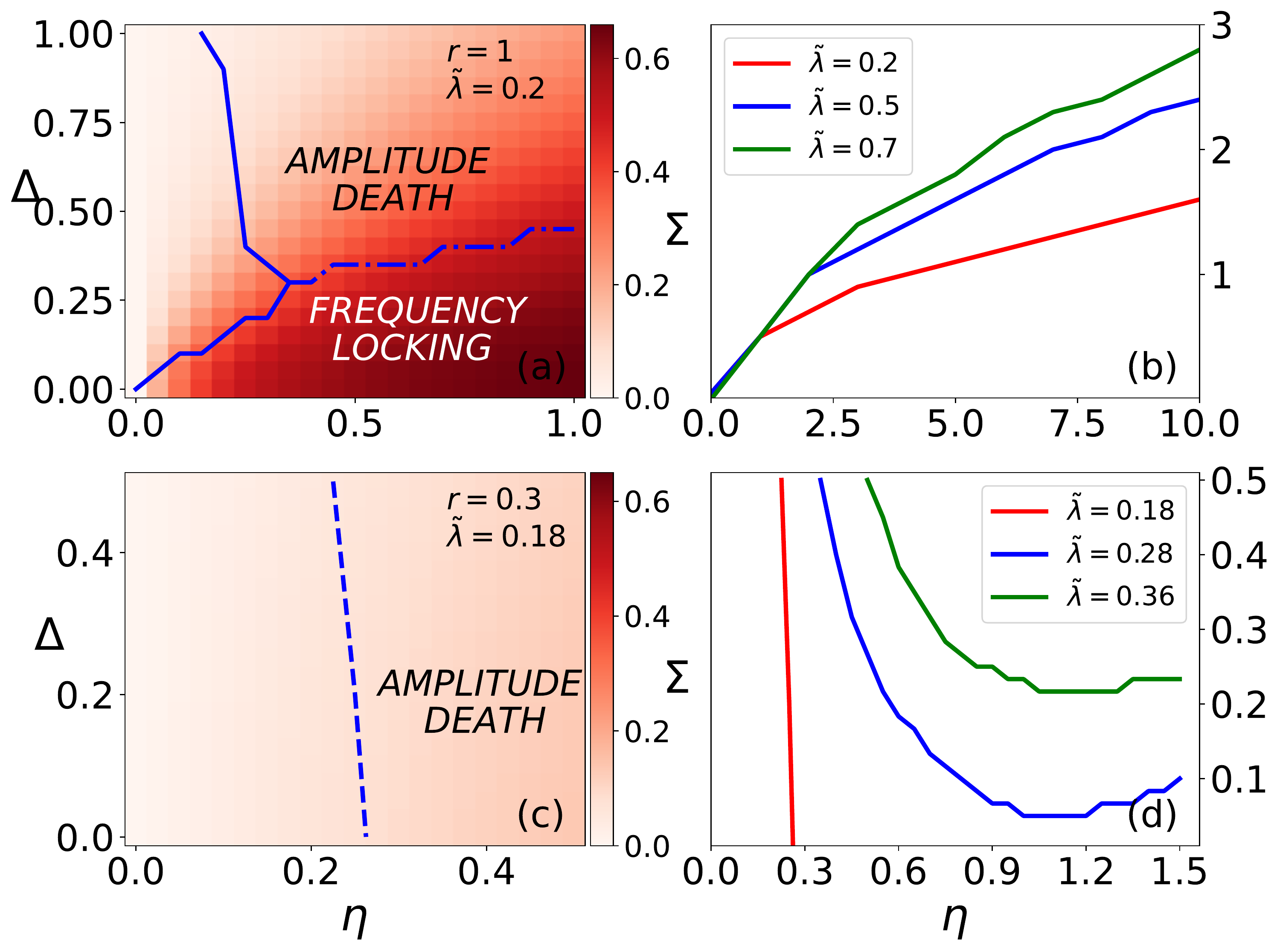}
\caption{Regions of frequency locking and amplitude death (as defined in the text by the power spectrum and Wigner function) for two dissipatively coupled vdP oscillators [subplots (a)--(d)] along with contours of $\Sigma$ [subplots (a) and (c)]. All subplots have $\Delta$ on the vertical axis, and $\eta$ on horizontal axis which we also indicate in subplot (c). (a) Large $r$ (semiclassical regime). Note the region on the left does not correspond to any identifiable effect and is demarcated using a solid line while the boundary between frequency locking and amplitude death is a Hopf bifurcation, which we denote by a dash-dotted line. As $r$ is increased, the classical boundary is recovered. (b) Effect of varying $\lambda$ on synchronization for $r=1$. Boundaries of the frequency-locking region and its corresponding $\lambda$ are shown (i.e.~the frequency-locking region is the area underneath each curve). Increasing $\lambda$ can enlarge the synchronization bandwidth and induce frequency locking from a state of amplitude death. (c) Small $r$ (deep quantum regime). At small $r$, amplitude death occurs even at zero initial detuning, which is a quantum effect. Note that $\bar{\lambda}$ in (a) and (c) are approximately equal, leaving the differences between the two plots only as a result of quantum effects. (d) Shifts in amplitude-death boundary as $\bar{\lambda}$ is increased (quantum-to-semiclassical transition).} 
\label{fig:2osc_detuning}
\end{figure}

In Fig.~\ref{fig:2osc_detuning} we work out regions of frequency locking and amplitude death in the $(\eta,\Delta)$ parameter space for \eqref{eq:coupled duffing_vdp me}, along with $\Sigma$, shown as a contour in Fig.~\ref{fig:2osc_detuning}(a) and (c). Two especially interesting scenarios are---when the limit cycles are relatively large compared to quantum noise [Fig.~\ref{fig:2osc_detuning}(a)]; and when they become small, being more susceptible to quantum noise [Fig.~\ref{fig:2osc_detuning}(c)].

In Fig.~\ref{fig:2osc_detuning}(a) we have indicated the boundary between frequency locking and amplitude death by a dash-dotted line, while no identifiable phenomenon occurs to the left of the solid line. Note the dash-dotted line is a Hopf-bifurcation curve because the transition from amplitude death to frequency locking is facilitated by a Hopf bifurcation. Clearly, $\Sigma$ is larger inside the frequency-locking region. Especially significant here is the effect of the \vdp\ nonlinearity on synchronization. Whereas in the single-oscillator case the \vdp\ nonlinearity had only detrimental effects, it now has a constructive role by enlarging the (mutual) synchronization bandwidth. We illustrate this in Fig.~\ref{fig:2osc_detuning}(b) where additional frequency-locking boundaries for different values of \vdp\ nonlinearity $\lambda$ are plotted. It can be seen that increasing $\lambda$ enlarges the frequency-locking region (see also Appendix for some classical analysis). This means that two oscillators in a state of amplitude death with an $(\eta,\Delta)$ lying above the Hopf-bifurcation curve in Fig.~\ref{fig:2osc_detuning}(a) will transit suddenly to a state of synchronized oscillations when $\lambda$ is sufficiently increased. This may be appropriately called nonlinearity-induced mutual synchronization. 

Turning now to the case of a small limit cycle $(r\ll1)$ in Fig.~\ref{fig:2osc_detuning}(c), we find frequency locking to be absent while position correlations become negligible. The blue solid line delineates the boundary of amplitude death. Most striking is the persistence of amplitude death at zero detuning (i.e.~$\Delta=0$). Classically, some frequency mismatch between the two oscillators must be present in order for amplitude death to occur \cite{SPR12}. This signifies a clear distinction between classical and quantum dynamics. A loss of amplitude death at $\Delta=0$ may then be expected if we increased either $r$ or $\lambda$ (while holding the other constant). This is indeed what we find, as illustrated in Fig.~\ref{fig:2osc_detuning}(d), where such a quantum-to-semiclassical transition is captured by increasing $\bar{\lambda}\equiv\lambda r^2$. 

Since we have focused exclusively on the \vdp\ nonlinearity here, we note that incorporating a Duffing nonlinearity into our model will not in fact change the frequency-locking boundary. We can understand this from the classical coupled equations of motion, where we see that $\beta$ appears only in the phase dynamics of the oscillators, and that such terms vanish at steady state for identical oscillators (equal limit cycle radii) \cite{aronson1990amplitude}.


\begin{figure}[t]
    \centering
    \includegraphics[width=\linewidth]{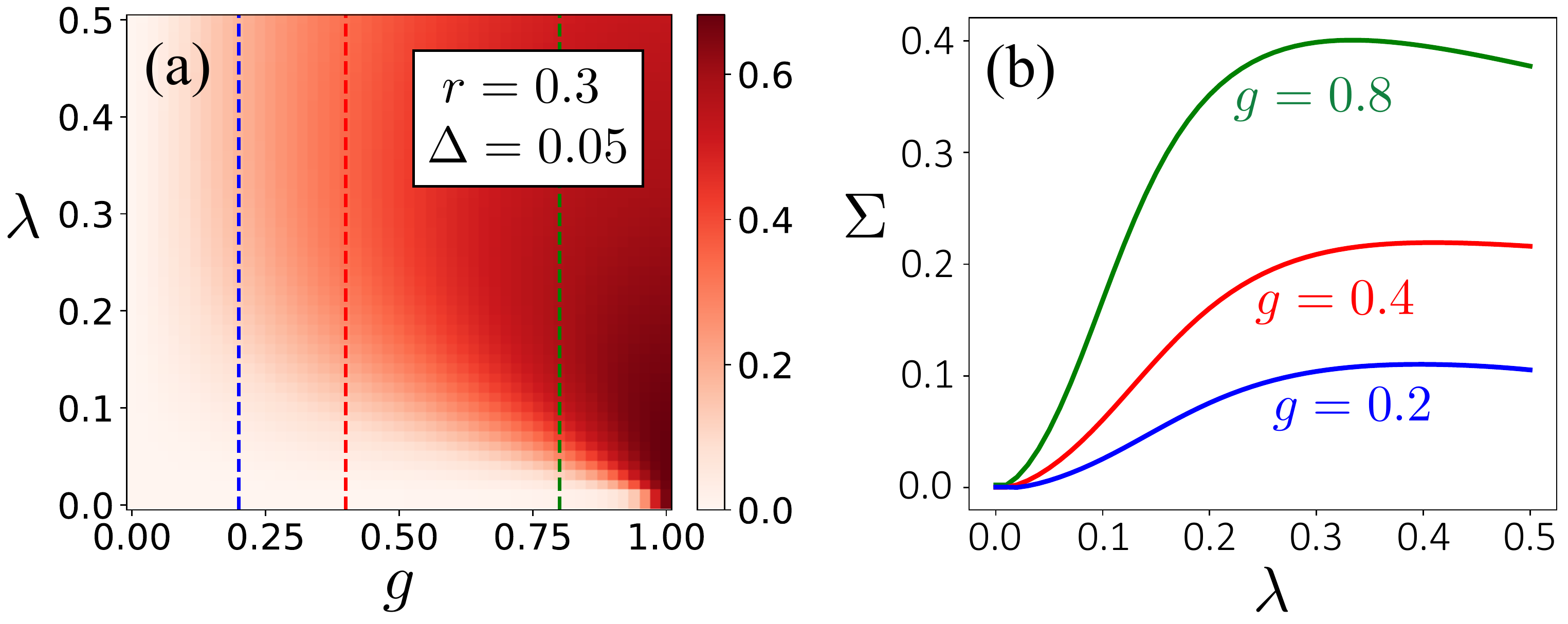}
    \caption{Positional correlation \eqref{PearCorr} for two reactively coupled \vdp\ oscillators. (a) Contour of $\Sigma$ as a function of $\lambda$ and $g$ for $r=0.3$ and $\Delta=0.05$. The bottom gap of zero correlation agrees with the \ls\ limit. (b) Correlations along the three vertical dashed lines at $g=0.2$, $0.4$, and $0.8$ in subplot (a).}
    \label{fig:full_vdp reactive coupled}
\end{figure}

{\it Nonlinearity-induced correlations.---}It is known that two reactively coupled \ls\ oscillators cannot synchronize nor share position correlations. This is true even in the quantum case~(see Appendix). However, we show here that positional correlations between two reactively coupled quantum \vdp\ oscillators do develop. Here we must use the exact \vdp\ Lindbladian~\cite{chia2020relaxation}, because under reactive coupling, the approximate model does not produce any off-diagonal elements in the steady state, and hence cannot generate correlations in the two oscillators. As with the dissipatively coupled system, two reactively coupled \vdp\ oscillators can be modelled by considering two uncoupled \vdp\ oscillators with annihilation operators $\ann_1$ and $\ann_2$, coupled by the Hamiltonian $g(\ann_1 \adg_2 + \adg_1 \ann_2)$, where $g$ is the reactive coupling strength. As before, we assume that both oscillators have the same nonlinearity $\lambda$.

In Fig.~\ref{fig:full_vdp reactive coupled}(a), we generate a contour plot of $\Sigma$ as a function of $\lambda$ and $g$ at $r=0.3$ and $\Delta=0.05$. From this we see that for a fixed $g$, increasing the oscillator nonlinearity beyond the \ls\ regime leads to stronger correlations. We illustrate this more clearly in Fig.~\ref{fig:full_vdp reactive coupled}(b) by showing how $\Sigma$ varies as a function of $\lambda$ for $g=0.2$, 0.4, and 0.8, which are marked in Fig.~\ref{fig:full_vdp reactive coupled}(a) by vertical dashed lines. Note this also shows the existence of an optimal $\lambda$ which maximizes $\Sigma$. Such nonlinearity-induced correlations are absent in the corresponding classical model. At a given coupling strength, the position correlation in two reactively coupled classical \vdp\ oscillators decreases monotonically as $\lambda$ increases~(see Appendix).

\textit{Conclusion.---}
Our work goes beyond the well-studied paradigm of weak nonlinearity in quantum synchronization, and provides the first systematic study of quantum synchronization effects for strong nonlinearity. We introduced a new quantum oscillator model which captures intriguing effects induced by two strong nonlinearities. We showed that a strong Duffing nonlinearity leads to a linear enhancement of the synchronization bandwidth in driven oscillators. We also reported genuine quantum synchronization effects exclusive to strong nonlinearity which are not observed previously: Increasing the vdP nonlinearity enhances the synchronization bandwidth, and revives synchronization between dissipatively-coupled oscillators in amplitude death. For reactively-coupled vdP oscillators on the other hand, we find that strong nonlinearity induces position correlations which are impossible in the weakly-nonlinear limit. Our model provides a new paradigm for studying other strongly nonlinear effects such as chaos~\cite{Sto99,Str15,Haa00,Bra01}.

\begin{acknowledgments}
\textit{Acknowledgements.---}YS and WJF would like to thank the support from NRF-CRP19-2017-01. CN was supported by the National Research Foundation of Korea (NRF) grant funded by the Korea government (MSIT) (NRF-2022R1F1A1063053). WKM, AC and LCK are grateful to the National Research Foundation, Singapore and the Ministry of Education, Singapore for financial support. 
\end{acknowledgments}

\appendix

\section{Exact quantum model}
Here, we describe how the quantum master equation for the exact Duffing-van der Pol model is obtained, following the approach in~\cite{chia2020relaxation}. Starting from the classical equation (2) in the main text, and defining the complex amplitude $\alpha = (\tilde{x} + i \tilde{y})/2$, we get the equation of motion
\begin{multline}
\label{eq:full_vdp_complex}
    \alpha' = i\,\frac{F}{2} \cos(\omega_{\rm d} t) -i\,\alpha - i\,\frac{\beta}{2} (\alpha + \alpha^*)^3 \\
    - \frac{\lambda}{2} (\alpha^3 + |\alpha|^2 \alpha - |\alpha|^2 \alpha^* - \alpha^{*3} - r^2 \alpha + r^2 \alpha^*) \,.
\end{multline}

Quantum theory defines the state of a dynamical system by a density operator $\rho$ satisfying $\rho'=\Lcal\rho$, where $\Lcal$ is a linear superoperator. By regarding $(\ann,\adg)$ as the analog of $(\alpha,\alpha^*)$, where $[\ann,\adg]=\hat{1}$, we can quantize a system prescribed by $\alpha'=h(\alpha,\alpha^*)$ by searching for an $\Lcal$ such that in the Schr\"{o}dinger picture, $\langle\ann\rangle'={\rm Tr}[\ann\,\Lcal\rho] =\langle:\!\!h(\ann,\adg)\!\!:\rangle$. Note that we have chosen to normally order $h(\ann,\adg)$, denoted by colons [e.g.~if $h(\ann,\adg)=\ann\adg$, then $:\!h(\ann,\adg)\!:\,=\adg\ann$]. Such an $\Lcal$ corresponding to \eqref{eq:full_vdp_complex} can be found in Lindblad form \cite{Lin76,GKS76,BP02}, and which we refer to as a Lindbladian:  
\begin{equation}
\label{eq:full duffing_vdp me}
    \Lcal = -i\,[\Hhat,\supopdot] 
                + \lambda \Dcal[\ann^\dagger \ann - \ann^{\dagger2}/2 ]  \nn + \lambda r^2 \Dcal[\ann^\dag] + \frac{3 \lambda}{4} \Dcal[\ann^2]  \, , 
\end{equation}
where
\begin{multline}
\label{eq:full duffing_vdp hamiltonian}
    \Hhat ={} \adg \ann - \frac{F}{2} \cos(\omega_{\rm d} t) (\ann + \adg) + \frac{3\beta}{4} \adg{}^2 \ann^2 \\
    + \frac{\beta}{2} (\adg \ann^3 + \adg{}^3 \ann) + \frac{\beta}{8}(\ann^4 + \adg{}^4)   
     - i\frac{\lambda}{2}(\ann^2-\adg{}^2) \\ - i \frac{\lambda}{4} (\adg \ann^3 - \adg{}^3 \ann) - i \frac{\lambda}{8}(\ann^4-\adg{}^4)  \,.
\end{multline}
We remark that the choice of the master equation is not unique, since we only demand that the classical equation of motion is recovered in the mean-field limit. Different choices of the master equation will in general lead to different quantum noise effects, which are most pronounced at low excitations (sometimes referred to as the deep quantum limit~\cite{mok2020/physrevresearch.2.033422}).


\section{A single forced classical oscillator}

\subsection{Poincar\'e--Lindstedt method}

The Poincar\'e--Lindstedt method is a perturbative technique to find approximate periodic solutions~\cite{Str15}. Regular perturbation theory fails at long time due to the existence secular terms. Taking $\lambda$ to be the perturbative parameter, such secular terms in the van~der~Pol (vdP) oscillator are of the form $t \cos t$, which are unbounded in $t$. The Poincar\'e--Lindstedt method overcomes this problem by removing secular terms explicitly.

\begin{figure*}
\centering
\includegraphics[width=0.8\linewidth]{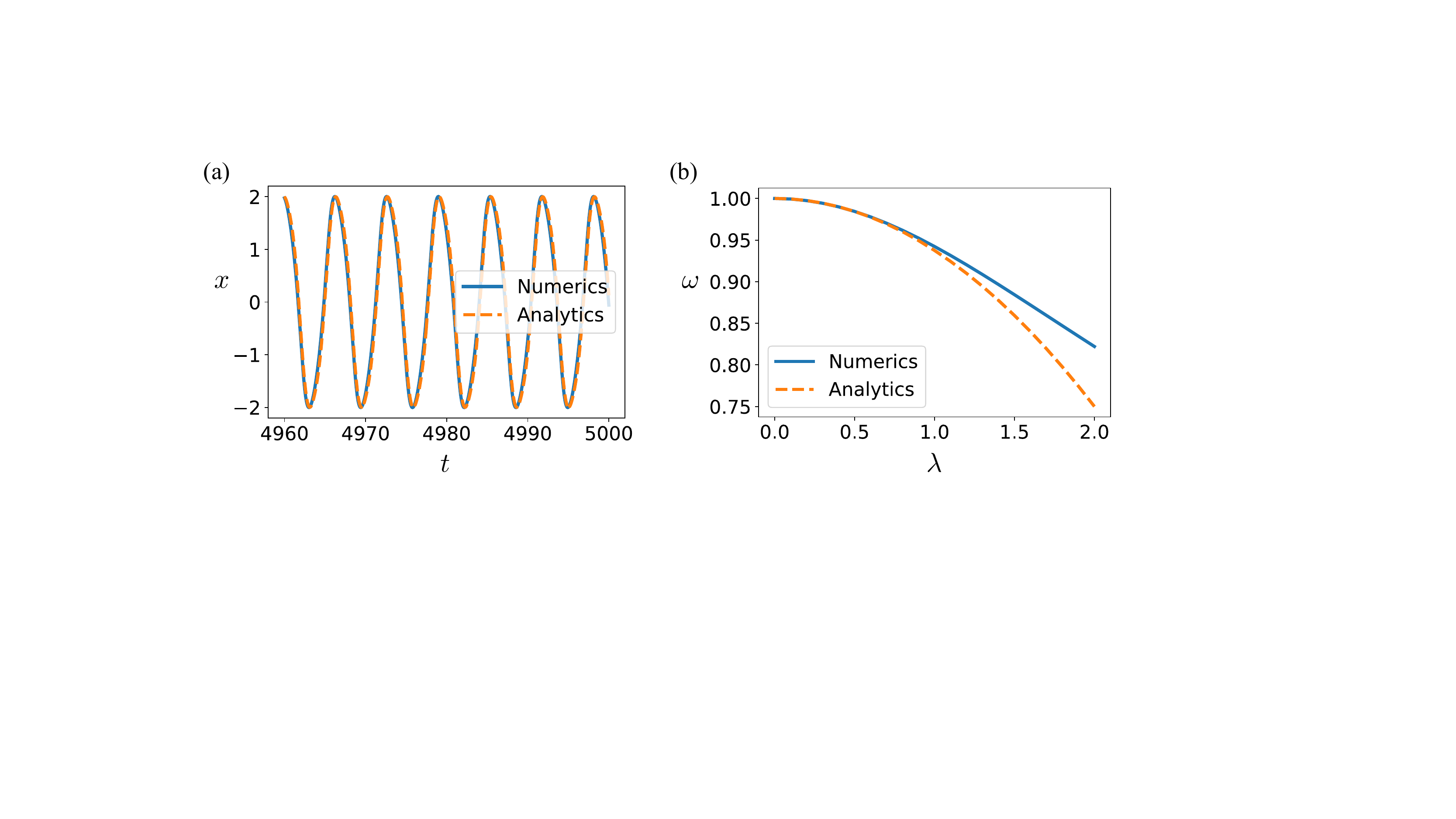}
	\caption{Dynamics of the undriven vdP oscillator. Results from an exact numerical simulation are shown as a blue solid line, while results from \eqref{eq:poincare_solution} are plotted as an orange dashed curve. (a) Long-time limit of $x(t)$ for $\lambda = 0.5$ (transient dynamics have been discarded). (b) Observed frequency $\omega$. The Poincar\'{e}--Lindstedt method remains to be a good approximation up till about $\lambda\approx1$.}
	\label{fig:undriven_vdP}
	\end{figure*}

Let us consider the undriven vdP oscillator (setting $F = 0$). Defining a rescaled time $\tau \equiv \omega t$ where $\omega$ is the oscillation frequency, we have the equation
\begin{equation}
    \omega^2 \, x''(\tau) + \lambda (x^2 - 1) \, \omega \, x'(\tau) + x(\tau) = 0  \; .
\end{equation}
where a prime denotes differentiation with respect to the argument. Now, we perform the perturbative expansions 
\begin{align}
    x(\tau) &= \breve{x}_0 (\tau) + \lambda \, \breve{x}_1(\tau) + \lambda^2 \, \breve{x}_2(\tau) + \rm{O}(\lambda^3) \; ,  \\
    \omega &= 1 + \lambda \, \breve{\omega}_1 + \lambda^2 \, \breve{\omega}_2 + \rm{O}(\lambda^3)  \; .
\end{align}  
Collecting terms of the same order in $\lambda$, the zeroth-order equation reads 
\begin{equation}
    \breve{x}''_0 + \breve{x}_0 = 0  \;,
\end{equation}
which just describes simple harmonic oscillations given by $\breve{x}_0(\tau) = a \cos \tau$, where $a=\breve{x}_0(0)$. The first-order equation reads
\begin{align}
    \breve{x}''_1 + \breve{x}_1 
    = {}& - 2 \, \breve{\omega}_1 \breve{x}''_0 - (\breve{x}_0^2 - 1) \breve{x}''_0   \nn \\
    = {}& \underbrace{2 \, \breve{\omega}_1 a \cos \tau - a \left(1 - \frac{a^2}{4}\right) \sin \tau}_{\text{resonant forcing}} + \frac{a^3}{4} \,  \sin(3\tau)  \;.
\end{align}
The resonant forcing gives rise to secular terms such as $\tau \cos \tau$ and $\tau \sin \tau$ which are unwanted. Hence, setting the resonant forcing to zero yields $\breve{\omega}_1 = 0, a = 2$ which gives $\breve{x}_0(\tau) = 2 \cos \tau$, $\omega = 1 + \text{O}(\lambda^2)$. This is the Stuart--Landau (SL), i.e.~weakly nonlinear limit of the vdP which describes quasiharmonic oscillations. Solving the resulting first-order equation, we get the leading-order correction for $x(\tau)$: $x_1(\tau) = \sin^3 \tau$. To obtain the leading-order correction for $\omega$, we look at the second-order equation
\begin{equation}
\begin{split}
    \breve{x}''_2 + \breve{x}_2 
    &= -(\breve{\omega}_1^2 + 2 \, \breve{\omega}_2) \, \breve{x}''_0 - 2 \, \breve{\omega}_1 \, \breve{x}''_1 \\
    &- (\breve{x}_0^2 - 1)\left( \breve{x}'_1 + \breve{\omega}_1 \, \breve{x}'_0 \right) - 2 \, \breve{x}_0 \, \breve{x}_1 \, \breve{x}'_0  \\ 
    &= \left(4 \, \breve{\omega}_2 + \frac{1}{4} \right) \cos \tau +  \text{higher harmonics}  \; .
\,\end{split}
\end{equation}
Again, eliminating the resonant forcing, we get $\breve{\omega}_2 = -1/16$. In terms of $t$, we then have
\begin{align}
    x(t) &= 2 \cos(\omega t) + \lambda \sin^3(\omega t) + \rm{O}(\lambda^2) \; , \\ 
    \omega &= 1 - \frac{\lambda^2}{16} + \rm{O}(\lambda^3)  \; .
\label{eq:poincare_solution}
\end{align}
Thus, the effect of small nonlinearity on the limit cycle is a decrease in frequency and a distortion without a change
in the limit cycle amplitude (from the first-order approximation of $x(t)$ we can see that for $\lambda < 4/3$ the amplitude of the oscillation is constant at 2. Hence, the amplitude is unchanged to order $\lambda$.) A numerical simulation of $x(t)$ and observed frequency $\omega$ as given by~\eqref{eq:poincare_solution} are plotted in Fig. \ref{fig:undriven_vdP}. Good agreement between the analytic and numeric simulation can be seen for $\lambda < 1$.

\subsection{Krylov--Bogoliubov averaging}

The Krylov--Bogoliubov averaging method essentially assumes that the amplitude is slowly varying and can be treated as a constant when averaging the dynamics over one period, which produced the time-averaged equations of motion. Introducing the complex amplitude, 
\begin{align}
\label{alphaDefn}
    \alpha(t) = \frac{1}{2} \, [ \, x(t) + i \, y(t) \,]  \; ,
\end{align}
where $y=x'$, we can rewrite the vdP equation as a complex equation of motion
\begin{equation}
    \alpha' = -i\,\alpha + i\,\frac{F}{2} \cos(\omega_{\rm d} t) 
                   - \frac{\lambda}{2} \, (\alpha^3 - \alpha^{*}{}^3 + |\alpha|^2 \alpha - |\alpha|^2 \alpha^*  - \alpha + \alpha^*)  \; .
\end{equation}
Performing the Krylov--Bogoliubov time averaging to first-order in $\lambda$, we obtain
\begin{equation}
    \alpha' = -i \, \alpha + \frac{\lambda}{2} \, (1-|\alpha|^2) \, \alpha \; ,
\label{eq:1storderavg}
\end{equation}
which is the well-known SL equation representing the normal form of a supercritical Hopf bifurcation. The stable oscillations are described by $\alpha(t) = \exp(-it)$. This gives $x(t) = 2 \cos t$ which is consistent with the Poincar\'e--Lindstedt method up to zeroth order in $\lambda$.

The second-order averaging yields~\cite{sanders2007averaging}
\begin{equation}
    \alpha' = -i \, \alpha + \frac{\lambda}{2} (1 - |\alpha|^2) \, \alpha + i \, \frac{\lambda^2}{8} \, \left( 1 - 6 \, |\alpha|^2 + \frac{11}{2} \, |\alpha|^4 \right) \alpha \; ,
\label{eq:2ndorderavg}
\end{equation}
which predicts a limit cycle amplitude of $\breve{x}_0 = 2|\alpha| = 2$ and frequency $\omega = 1 - \lambda^2 / 16$, again consistent with the zeroth-order Poincar\'e--Lindstedt method.

\subsection{Synchronization analysis}
Here, we use the harmonic-balance method to analyze the synchronization behavior for the Duffing--van~der~Pol (DvdP) equation in nondimensionalized form, given by
\begin{equation}
    x'' + \lambda \, ( x^2 - r^2 )x' + x + \beta x^3 = F \cos(\omega_{\rm d} t)  \; .
\label{eq:duffingvdP}
\end{equation}
Note that for ease of writing we have omitted tildes for $x$ and all dimensionless parameters (e.g.~time). We then assume a synchronized solution of the form $x(t) = A \cos (\omega_{\rm d} t - \phi)$, where $A$ is the amplitude of the motion and $\phi$ is a constant phase shift from the driving force. Substituting the ansatz for $x$ into \eqref{eq:duffingvdP}, we get
\begin{multline}
    -\omega_{\rm d}^2 A \cos(\omega_{\rm d} t - \phi) - \lambda \, \omega_{\rm d} A^3 \cos^2(\omega_{\rm d} t - \phi)\sin(\omega_{\rm d} t - \phi) \\
    + \lambda \,\omega_{\rm d} r^2 A \sin(\omega_{\rm d} t - \phi) + A \cos(\omega_{\rm d} t - \phi) \\
    + \beta A^3 \cos^3(\omega_{\rm d} t - \phi)= F \cos(\omega_{\rm d} t) \,.
\end{multline}

The second term on the left-hand side may be written as
\begin{multline}
    \cos^2(\omega_{\rm d} t - \phi) \sin(\omega_{\rm d} t - \phi) = \frac{1}{4}\sin(\omega_{\rm d} t - \phi) \\
    + \text{higher harmonics}  \; .
\end{multline}
Neglecting the higher harmonics and then collecting the coefficients of $\cos(\omega_{\rm d} t)$ and $\sin(\omega_{\rm d} t)$, we obtain
\begin{multline}
    (1 - \omega_{\rm d}^2) A \cos \phi + \frac{1}{4} \lambda \, \omega_{\rm d} A^3 \sin \phi - \lambda \, \omega_{\rm d} A r^2 \sin\phi + \frac{3}{4} \beta A^3 \cos \phi \\
    = F  \; ,\\
    (1 - \omega_{\rm d}^2) A \sin \phi - \frac{1}{4} \lambda \, \omega_{\rm d} A^3 \cos \phi + \lambda \, \omega_{\rm d} A r^2 \cos\phi + \frac{3}{4} \beta A^3 \sin \phi \\
    = 0  \; .
\end{multline}
These may be expressed compactly as a single complex equation as
\begin{equation}
    \big(1-\omega_{\rm d}^2\big)A - i \lambda \, \omega_{\rm d} A \left( \frac{A^2}{4}  - r^2 \right) + \frac{3}{4} \, \beta A^3 = F e^{-i\phi}  \; .
\label{eq:duffingvdP_complex}
\end{equation}
\begin{align}
\label{PertExpansion}
    A = A_\star + p \,\delta \!\!\;A  \;,  \quad 
    \omega_{\rm d} = \omega_\star + p \, \delta\omega  \; .
\end{align}
Substituting \eqref{PertExpansion} into \eqref{eq:duffingvdP_complex} then gives the first-order equation in $p$,
\begin{equation}
    \delta \!\!\;A \left( 6 \beta r^2 - 2 i \sqrt{1 + 3 \beta r^2} \, \lambda r^2 \right) - 4 r \sqrt{1 + 3 \beta r^2} \, \delta \omega = \lambda \, e^{-i \phi}  \; .
\end{equation}
The modulus of the left-hand side must be $\lambda$, hence we have
\begin{multline}
    \left(6 \beta r^2 \delta\!\!\;A - 4r \sqrt{1 + 3 \beta r^2}\, \delta \omega\right)^2 + \left(2 \lambda r^2 \sqrt{1 + 3\beta r^2} \, \delta\!\!\;A\right)^2 \\
    = \lambda^2  \; .
\label{eq:duffingvdP_modulus}
\end{multline}
This can be solved to obtain the relationship between $\delta\!\!\;A$ and $\delta\omega$. In order to derive the synchronization frequency range, we first notice that \eqref{eq:duffingvdP_modulus} is a quadratic equation in $\delta\;\!\!A$. For synchronization to exist, $\delta\!\!\;A$ must be real, hence the discriminant must be non-negative, i.e.,
\begin{multline}   
    - 4 \left\{4r^4 \left[ 9 \beta^2 + \lambda^2(1+3\beta r^2) \right] \right\} \left[ 16 \, r^2 (1+3\beta r^2) \, \delta\omega^2 - \lambda^2 \right] \\
    +\left( 48 \beta r^3 \sqrt{1+3\beta r^2} \, \delta \omega \right)^2 \geq 0  \; ,
\end{multline}
which yields
\begin{equation}
    \delta\omega^2 \leq \frac{9\beta^2 + \lambda^2 (1 + 3 \beta r^2)}{16r^2(1+3\beta r^2)^2} \equiv \omega_{\rm c}^2  \; .
\end{equation}
Hence, the vdP oscillator is synchronized if the driving frequency is within the critical interval
\begin{equation}
    \omega_0 - \frac{F}{\lambda} \omega_{\rm c} \leq \omega_{\rm d} \leq \omega_0 + \frac{F}{\lambda} \omega_{\rm c}  \; .
\end{equation}
The synchronization bandwidth is thus
\begin{equation}
    \frac{2F}{\lambda} \, \omega_{\rm c} = \frac{(F/r)}{2\lambda r^2 (1+3\beta r^2)} \sqrt{(\lambda r^2)^2 (1+3\beta r^2) + 9(\beta r^2)^2}  \; .
\end{equation}
Evidently, by fixing $F/r$, $\lambda r^2$, and $\beta r^2$, the synchronization bandwidth becomes independent of the scale parameter $r$. As shown in the main text, this does not hold true in the quantum case due to quantum noise. Defining
\begin{align}
    \bar{F} = F/r \;,  \quad \bar{\lambda} = \lambda r^2 \;,  \quad \bar{\beta} = \beta r^2 \;,
\end{align}
the bandwidth can be rewritten as
\begin{align}
        \frac{2\bar{F}}{\bar{\lambda}} \omega_c &= \frac{\bar{F}}{2\bar{\lambda} (1+3\bar{\beta})} \sqrt{\bar{\lambda}^2 (1+3\bar{\beta}) + 9 \bar{\beta}^2} \nonumber\\
        &\approx \begin{cases} \bar{F}/2\bar{\lambda} \;, &\quad \bar{\beta} \longrightarrow \infty \\ \bar{F}/2 \;, &\quad \bar{\beta} \longrightarrow 0 \end{cases}
\end{align}
where in the last step we have taken the limit of large and small $\bar{\beta}$. For the usual vdP oscillator ($\bar{\beta}=0$), the synchronization bandwidth reduces to $\bar{F}/2$ which is independent of $\lambda$. In order to obtain nonlinearity-induced enhancement in the bandwidth, we require
\begin{equation}
    \frac{\bar{F}}{2\bar{\lambda} (1+3\bar{\beta})} \, \sqrt{\bar{\lambda}^2 (1+3\bar{\beta}) + 9 \bar{\beta}^2} > \frac{\bar{F}}{2}  \; .
\end{equation}
This results in the conditions
\begin{equation}
    \bar{\beta} > \frac{\bar{\lambda}^2}{3(1- \bar{\lambda}^2)} \;,  \quad 0 < \bar{\lambda} < 1   \;.
\end{equation}
Without loss of generality, we can set $r = 1$ in the following discussion. In this case $\bar{\lambda}=\lambda$, $\bar{\beta}=\beta$, and $\bar{F}=F$. Interestingly, the nonlinearity-induced enhancement only acts for a finite range of $\lambda$ and requires a minimum $\beta$. In the limit of large $\beta$, the bandwidth tends to $F / 2\lambda$. However, the method of harmonic balance assumes $\lambda$ and $\beta$ to be weak. In this regime, we can interpret the existence of a minimum $\beta$ as a competition between the effects of $\lambda$ and $\beta$. Note that the oscillator frequency is $\omega \approx 1 + 3\beta/2 - \lambda^2/16$ from our previous analysis. The threshold for synchronization enhancement occurs when these two opposite effects on the frequency are of the same scale, i.e. $\beta \sim \lambda^2$.
\begin{figure*}
\centering
  \includegraphics[width=0.8\linewidth]{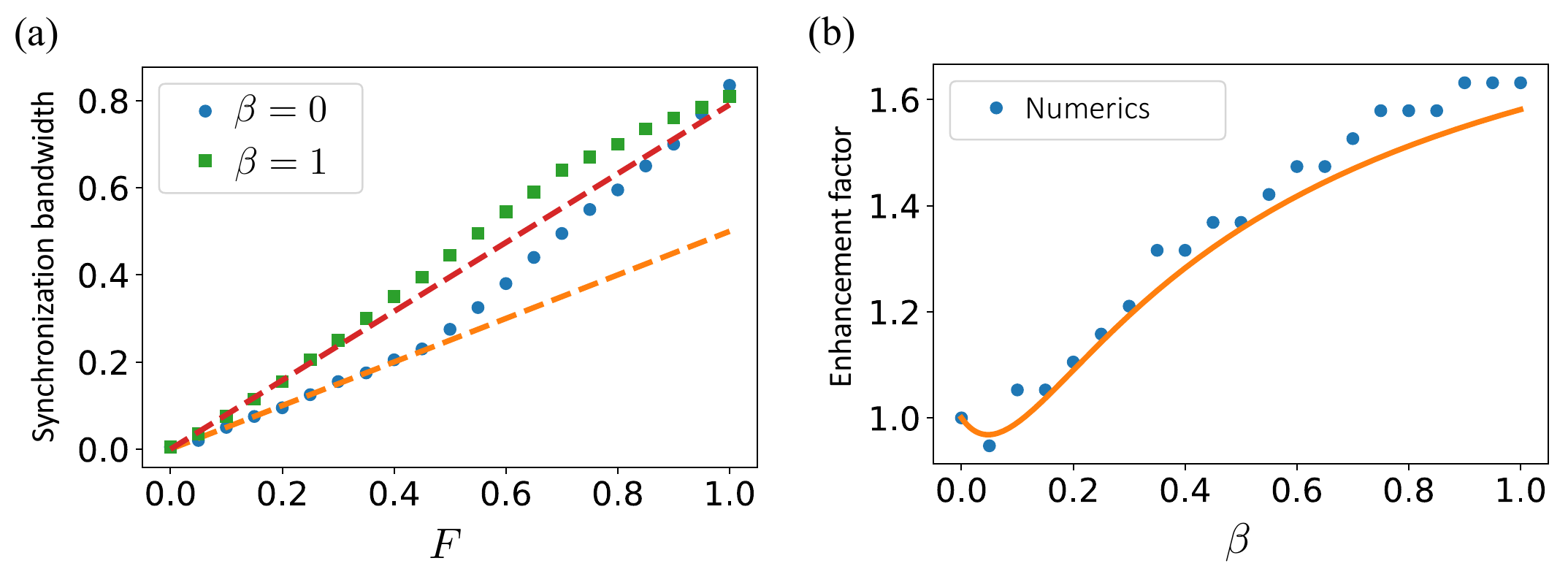}
	\caption{(a) Synchronization bandwidth against driving force $F$ for $\lambda = 0.5$. Adding $\beta$ enhances the synchronization bandwidth for weak to moderate forcing (compared to $\lambda$). The analytical predictions given by the dashed lines) are accurate up to $F \sim \lambda$. (b) Enhancement factor defined as the ratio between the bandwidth and its baseline value $F/2$. A value greater than one indicates enhancement of the bandwidth. $\lambda = 0.5, F = 0.2$, and the predicted minimum $\beta$ for enhancement is $1/9$.}
\label{fig:vdP_enhancement}
	\end{figure*}

Figure~\ref{fig:vdP_enhancement} shows the effect of adding the Duffing nonlinearity on the synchronization bandwidth for $r=1$. Comparing $\beta = 0$ and $\beta = 1$ in Fig.~\ref{fig:vdP_enhancement}(a), we can see that the synchronization bandwidth is enhanced for the $\beta = 1$ case for weak to moderate forcing $F$ (compared to $\lambda$). The analytical prediction of the bandwidth agrees with the numerical simulations up to $F \sim \lambda$, beyond which the numerical bandwidth exceeds the predicted value indicating the suppression of natural dynamics at strong forcing. Using $\lambda = 0.5$, we also predict that the synchronization enhancement occurs when $\beta > 1/9$. Indeed, by plotting in Fig.~\ref{fig:vdP_enhancement}(b) the enhancement factor, defined as the ratio between the bandwidth and its baseline value $F/2$, we see a good agreement between the numerical results and the analytical calculations. Fluctuations in the numerics can be attributed to a few reasons, such as the time resolution $dt$ in the simulation, the number of samples used for $x(t)$, the frequency resolution $d\omega$ used in determining the bandwidth, and the threshold observed detuning used to determine the frequency-locked regions (set to 0.001). The numerical results also appear to deviate slightly from the prediction at large $\beta$, which is likely due to the breakdown in making the ansatz $x(t)=A\cos(\omega_{\rm d}\,t-\phi)$ used in the derivation. Including higher harmonics in the ansatz might result in more accurate predictions at larger $\beta$.


\section{Two classical oscillators with dissipative coupling}

The equations of motion for two dissipatively coupled vdP oscillators are often written as
\begin{equation}
\begin{split}
    x''_1 + \lambda(x_1^2 - 1) \, x'_1 + x_1 = \frac{\eta}{2} \, (x'_2 - x'_1) \; , \\  
    x''_2 + \lambda(x_2^2 - 1) \, x'_2 + (1+\Delta)x_2 = \frac{\eta}{2} \, (x'_1 - x'_2) \; ,  
\label{eq:2vdP}
\end{split}
\end{equation}
where $\eta$ is the coupling strength and $\Delta$ is the detuning between the two oscillators. The system has four phase-space dimensions, two for each oscillator, given by $(x_1,y_1)=(x_1,x'_1)$ and $(x_2,y_2)=(x_2,x'_2)$. We can express the coupled system more simply using complex amplitudes as we did in \eqref{alphaDefn}, except now
\begin{align}
    \alpha_k(t) = \frac{1}{2} \, [ \, x_k(t) + i \, y_k(t) \, ]  \;,  \quad k=1,2 \;.
\end{align} 
We will analyze the following first-order averaged equations for the coupled system assuming the nonlinearity to be weak,
\begin{equation}
\begin{split}
    \alpha'_1 &= -i \,\alpha_1 + \frac{\lambda}{2} \, (1-|\alpha_1|^2)\alpha_1 + \frac{\eta}{2} \, (\alpha_2 - \alpha_1)  \;,  \\ 
    \alpha'_2 &= -i\,(1+\Delta) \alpha_2 + \frac{\lambda}{2} \, (1-|\alpha_2|^2)\alpha_2 + \frac{\eta}{2} \, (\alpha_1 - \alpha_2) \;.
\end{split}    
\end{equation}

\subsection{Synchronization analysis}
\label{vdPSyncAnalysis}
In fact, only three real variables are needed in polar coordinates. If we write %
\begin{align}
    \alpha_k = R_k \exp{(-i \phi_k)} \;, \quad  k=1,2, 
\end{align}
then we may express the coupled dynamics in terms of only $R_1$, $R_2$, and the phase difference $\varphi \equiv \phi_2 - \phi_1$:
\begin{equation}
\begin{split}
    R'_1 &= \frac{\lambda}{2} \, R_1 \, (1-R_1^2) + \frac{\eta}{2} \, (R_2 \cos \varphi - R_1) \;, \\ 
    R'_2 &= \frac{\lambda}{2} \, R_2 \, (1-R_2^2) + \frac{\eta}{2} \, (R_1 \cos \varphi - R_2) \;, \\ 
    \varphi' &= \Delta - \frac{\eta}{2} \,\left( \frac{R_1}{R_2} + \frac{R_2}{R_1} \right) \sin \varphi  \;.
\end{split}
\end{equation}
By symmetry, if the two oscillators synchronize, we must have $R_1 = R_2 = R$. In this case the above equations simplify further to
\begin{equation}
\begin{split}
    R' &= \frac{\lambda}{2} \, R (1-R^2) + \frac{\eta}{2} \, R (\cos \varphi - 1) \;,  \\ 
    \varphi' &= \Delta - \eta \sin \varphi  \;.
\end{split}
\label{eq:twovdP_symmetrical}
\end{equation}
In the synchronized state, $R'=\varphi'=0$. Solving the phase equation, we get two solutions, $\varphi_\star^{(1)} = \sin^{-1}(\Delta/\eta)$ and $\varphi_\star^{(2)} = \pi - \sin^{-1}(\Delta/\eta)$. To determine the stable phase, we use linear stability analysis. Substituting $\varphi(t) = \varphi_\star^{(k)} + \epsilon(t)$ ($k = 1,2$) into the phase equation of motion and keeping only first-order terms in $\epsilon$, we get 
\begin{align}
    \epsilon' &= \Delta - \eta \,\sin\!\big[\varphi_\star^{(k)} + \epsilon\big] 
                   = \Delta - \eta \,\sin\!\left[ \sin^{-1}\! \bigg(\frac{\Delta}{\eta}\bigg) - (-1)^k \epsilon \right] 
                   \nonumber\\
                   &\approx (-1)^k \epsilon \, \eta \, \sqrt{1-\frac{\Delta^2}{\eta^2}}  \;.
\end{align}
From this we see that $\varphi_\star^{(1)}$ is stable while $\varphi_\star^{(2)}$ is unstable. We also find that $|\Delta| < \eta$ to be a necessary condition for synchronization. Substituting $\varphi_\star^{(1)}$ into the radial equation $R'= 0$ then gives
\begin{equation}
    \frac{\lambda}{2} \, R \, (1-R^2) + \frac{\eta}{2} \, R \left(\sqrt{1 - \frac{\Delta^2}{\eta^2}} - 1 \right) = 0 \;.
\end{equation}
This has the solution 
\begin{align}
    R^2_\star = 1 + \frac{\eta}{\lambda} \, \left( \sqrt{1 - \frac{\Delta^2}{\eta^2}} - 1 \right) \;. 
\end{align}
The zero-amplitude solution corresponds to the case of amplitude death and will be analyzed next. Performing linear stability analysis, we find that $r_\star$ is stable for $0<|\Delta| \leq \lambda$ and $|\Delta|<\eta$, or when, $|\Delta|>\lambda$ and $|\Delta|<\sqrt{\lambda(2\eta-\lambda)}$. Combining the phase and amplitude solutions, the conditions for synchronization is then
\begin{align}
   |\Delta| < \eta  \;, \quad \text{for $\eta \leq \lambda$}  \;,  \\
    |\Delta| < \sqrt{\lambda(2 \eta - \lambda)} \;,  \quad \text{for $\eta > \lambda$} \;. 
\label{eq:syncConditions}
\end{align}

\subsection{Amplitude death}

It is possible for coupled oscillators to achieve amplitude death, where the limit cycle oscillations are suppressed due to the mutual coupling. To obtain the conditions for amplitude death, we require the solution $\alpha_1 = \alpha_2 = 0$ to be stable. Since the phase is undefined in this case, we will analyze in Cartesian coordinates instead. Denoting $\alpha_k = \Re[\epsilon_k] + i\,\Im[\epsilon_k]$ ($k=1,2$) for small $\epsilon_k$, we can linearize the coupled equations around the origin, giving,

\begin{multline}
\label{TwoOscStabMatrix}
    \begin{pmatrix} \Re[\epsilon_1] \\ \Im[\epsilon_1] \\ \Re[\epsilon_2] \\ \Im[\epsilon_2]  \end{pmatrix}' 
    = \\
    \begin{pmatrix} (\lambda-\eta)/2 & \omega_1 & \eta/2 & 0 \\ -\omega_1 & (\lambda-\eta)/2 & 0 & \eta/2 \\ \eta/2 & 0 & (\lambda-\eta)/2 & \omega_2 \\ 0 & \eta/2 & -\omega_2 & (\lambda-\eta)/2 \end{pmatrix} 
    \begin{pmatrix} \Re[\epsilon_1] \\ \Im[\epsilon_1] \\ \Re[\epsilon_2] \\ \Im[\epsilon_2] \end{pmatrix}  \; ,
\end{multline}

where $\omega_1 = 1$ and $\omega_2 = 1 + \Delta$. Let $h$ be the eigenvalues of the stability matrix in \eqref{TwoOscStabMatrix}, and $p = h - (\lambda-\eta)/2$. The characteristic polynomial for $h$ then reads
\begin{multline}
    p^4 + p^2 \left( \omega_1^2 + \omega_2^2 - \frac{\eta^2}{2} \right) + \left( \frac{\eta^2}{4} + \omega_1 \, \omega_2 \right)^2 = 0 \; \\
    \implies\; \left[ p^2 - \left( \frac{\eta^2}{4} + \omega_1 \, \omega_2 \right) \right]^2 = -(\omega_1+\omega_2)^2 \, p^2  \;.
\end{multline}
The eigenvalues $h$ are therefore
\begin{equation}
    h = \frac{1}{2} \left[\lambda-\eta \pm \sqrt{\eta^2 - \Delta^2} \,\right] \pm i \, \frac{(\omega_1 + \omega_2)}{2}  \; .
\end{equation}
For $\alpha_1 = \alpha_2 = 0$ to be stable, the real part of $h$ must be negative. This gives us the the following necessary and sufficient condition for amplitude death
\begin{equation}
    \eta > \lambda \;, \quad |\Delta| > \sqrt{\lambda(2\eta - \lambda)}  \;.
\end{equation}
The curve $|\Delta| = \sqrt{\lambda\,(2\eta-\lambda)}$ thus separates the regions of synchronization and amplitude death, also known as the Hopf bifurcation curve.

\subsection{Effect of the Duffing nonlinearity}

Recall from \eqref{eq:duffingvdP} that a Duffing--van~der~Pol (DvdP) oscillator has a nonlinear frequency term $\beta x^3$ in its second-order equation of motion for $x$. If we were to define the DvdP by its equation of motion for its complex amplitude $\alpha$, then this term amounts to adding $-i3\beta|\alpha|^2\alpha$ to $\alpha'$ under first-order time averaging with respect to the Duffing nonlinearity. We show here that such a modification due to the Duffing nonlinearity in two dissipatively coupled DvdP oscillators does not change their mutual synchronization from the $\beta=0$ case. The classical time-averaged equations of motion for two dissipatively-coupled DvdP oscillators are
\begin{multline}
\label{DvdPOsc1}
    \alpha'_1 = -i \, \omega_1 \, \alpha_1 + \frac{\lambda}{2} \, (r^2 - |\alpha_1|^2)\alpha_1 \\
    + i \frac{\lambda^2}{8} \left( r^4 - 6r^2 |\alpha_1|^2 + \frac{11}{2}|\alpha_1|^4 \right)\alpha_1 \\
    - i \frac{3\beta}{2} \, |\alpha_1|^2 \alpha_1 + \frac{\eta}{2} \, ( \alpha_2 - \alpha_1 ) \;, 
\end{multline}
\begin{multline}
\label{DvdPOsc2}
    \alpha'_2 = -i \, \omega_2 \, \alpha_2 + \frac{\lambda}{2} \, (r^2 - |\alpha_2|^2)\alpha_2 \\
    + i \frac{\lambda^2}{8} \left( r^4 - 6r^2 |\alpha_2|^2 + \frac{11}{2}|\alpha_2|^4\right)\alpha_2 \\
    - i \frac{3\beta}{2} \, |\alpha_2|^2 \alpha_2 + \frac{\eta}{2} (\alpha_1 - \alpha_2) \;,
\end{multline} 
where $\omega_1 = 1$ and $\omega_2 = 1+\Delta$. Note these equations assume second-order time averaging with respect to the vdP nonlinearity [see \eqref{eq:2ndorderavg}]. As in the case of $\beta=0$ in Sec.~\ref{vdPSyncAnalysis}, mutual synchronization is more easily analyzed using polar coordinates, where $\alpha_k=R_k\exp(-i\phi_k)$ ($k=1,2$). Equations \eqref{DvdPOsc1} and \eqref{DvdPOsc2} are then equivalent to 
\begin{equation}
\begin{split}
    R'_1 &= \frac{\lambda}{2} \, \big( r-R_1^2 \big) R_1 + \frac{\eta}{2} \, \big(R_2 \cos\varphi - R_1\big) \;,  \\
    R'_2 &= \frac{\lambda}{2} \, \big( r- R_2^2 \big) R_2 + \frac{\eta}{2} \, \big(R_1 \cos\varphi - R_2\big) \;,  \\
    \varphi' &= \Delta - \frac{\lambda^2}{8} \left[ 6r^2\big(R_1^2 - R_2^2\big) - \frac{11}{2} \big(R_1^4 - R_2^4\big) \right] \\
    &- \frac{3\beta}{2} \big(R_1^2 - R_2^2\big) - \frac{\eta}{2} \left( \frac{R_1}{R_2} + \frac{R_2}{R_1} \right) \sin\varphi  \;.
\end{split}
\end{equation}
Recall from Sec.~\ref{vdPSyncAnalysis} that we have defined $\varphi \equiv \phi_2 - \phi_1$. By symmetry, when the two oscillators synchronize, we have $R_1 = R_2$. Consequently, the equation of motion for the phase difference $\varphi$ becomes independent of $R_1$ and $R_2$. Moreover, the $\beta$ term vanishes, which suggests that the $\beta$ nonlinearity has no effect on the mutual synchronization.

\subsection{Synchronization bandwidth}

\begin{figure}
\centering
\subfloat{%
  \includegraphics[width=\linewidth]{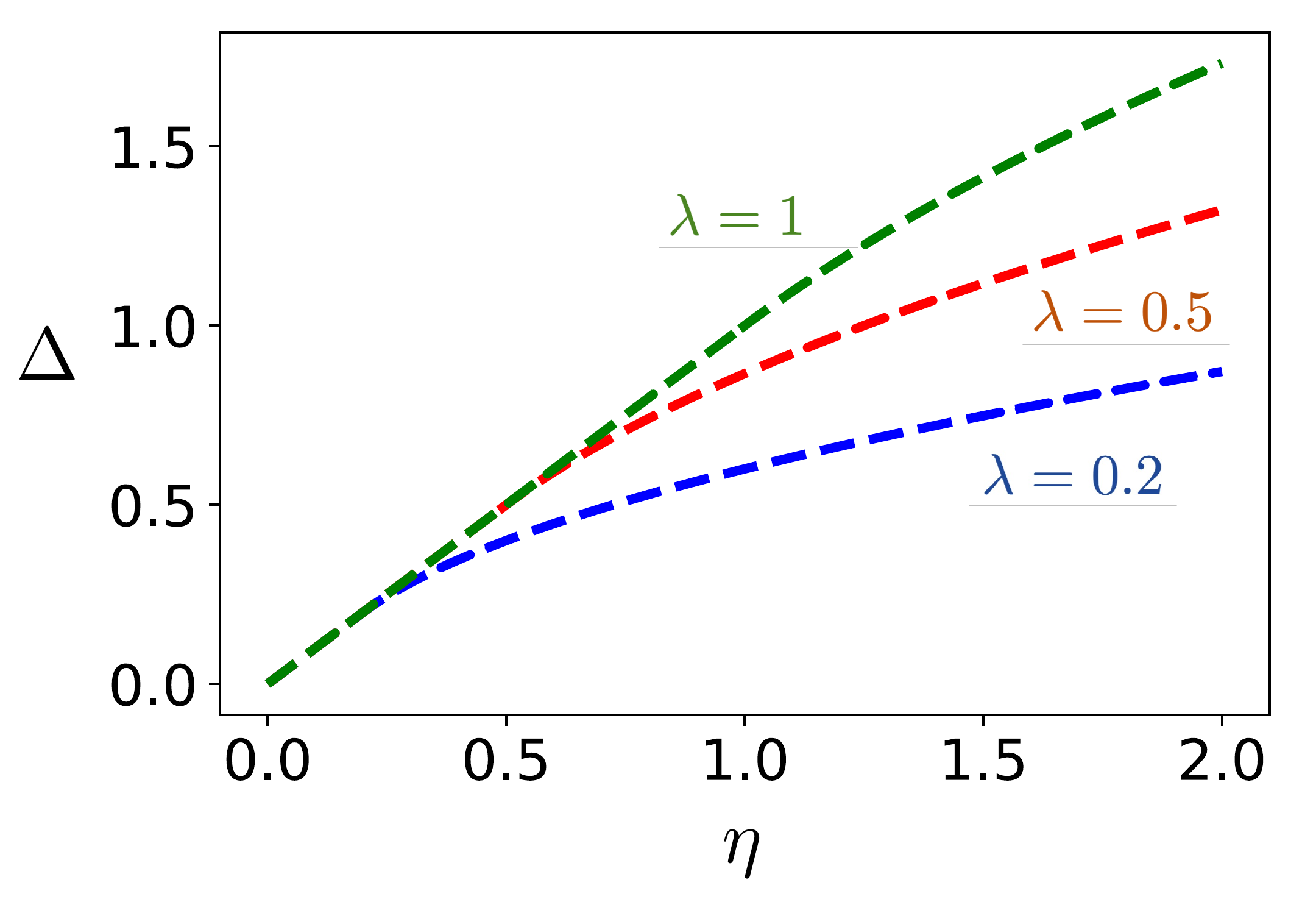}%
  \label{}%
}
	\caption{Synchronization boundaries for different values of $\lambda$. The synchronization region lies below the curve in the all cases. The synchronization bandwidth is enlarged as $\lambda$ is increased for a fixed $\eta$.}
	\label{fig:2vdP_boundary}
	\end{figure}

For a fixed $\eta$, we define from Eq.~\eqref{eq:syncConditions} the synchronization bandwidth to be the range of initial detunings $\Delta$ for which the two oscillators synchronize. This is given by the piecewise function
\begin{equation}
    \text{Bandwidth} = \begin{cases} 
      2\sqrt{\lambda(2\eta - \lambda)} & \lambda \leq \eta \\
      2\,\eta & \lambda > \eta
   \end{cases}
\end{equation}
where the factor of $2$ arises because $\Delta$ can be either positive or negative (or zero). Increasing $\lambda$ from 0, we find that the bandwidth increases monotonically until $\lambda = \eta$, after which the bandwidth saturates at $2\eta$. Thus, the vdP nonlinearity enlarges the synchronization bandwidth. This is shown by the synchronization boundaries in Fig. \ref{fig:2vdP_boundary} for various values of $\lambda$, where synchronization regions lie below the boundary curves. Fixing $\eta$ and increasing the nonlinearity parameter $\lambda$ leads to a larger synchronization bandwidth. This also implies that fixing $\Delta$ and $\eta$ while increasing $\lambda$ can lead to \textit{nonlinearity-induced} mutual synchronization whereby the two oscillators transition suddenly from amplitude death to synchronized oscillations.

\begin{figure}
\centering
\subfloat{%
  \includegraphics[width=\linewidth]{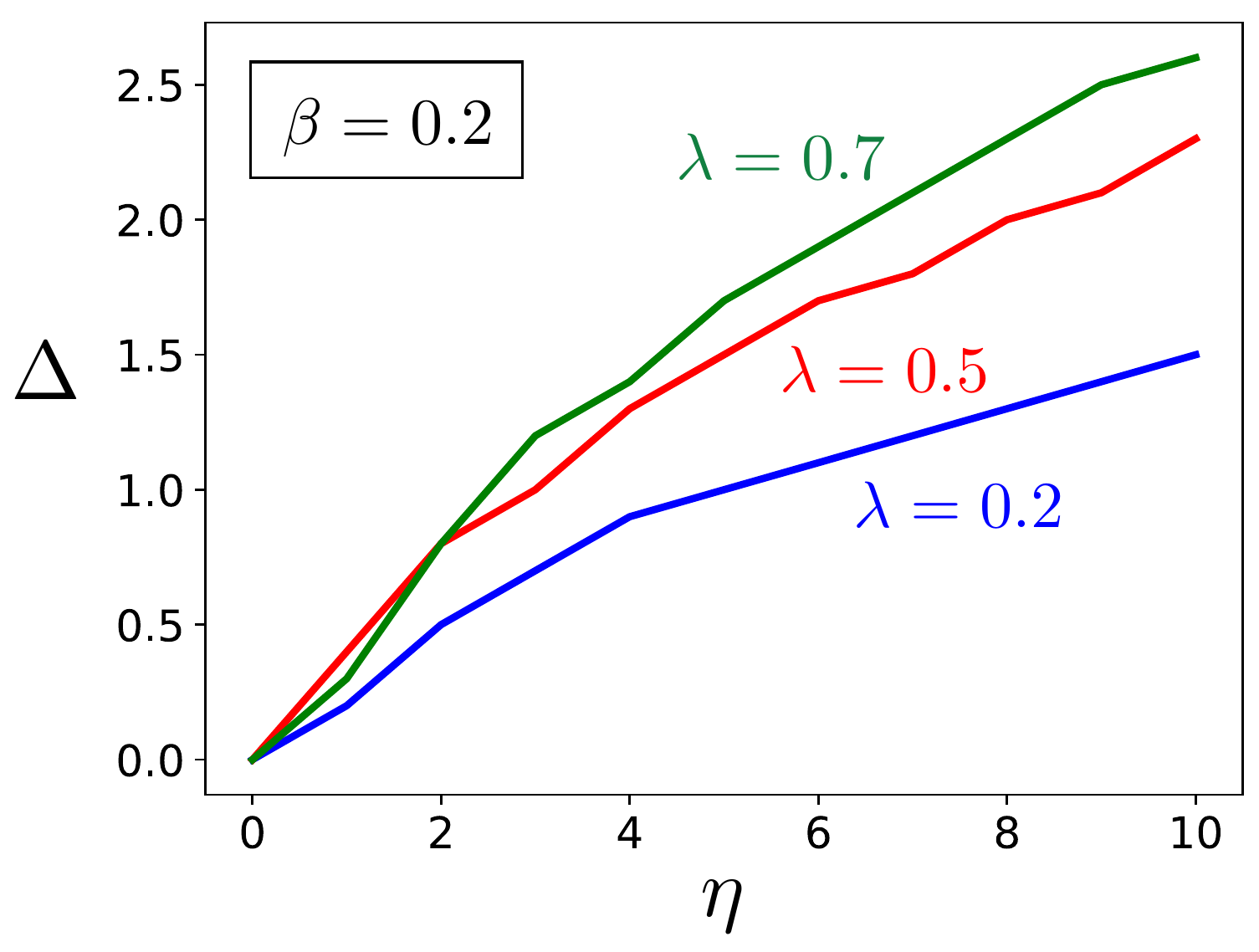}%
  \label{}%
}
\caption{Synchronization boundaries of two coupled quantum DvdP oscillators for different values of $\lambda$. Comparing with Fig.2b in the main text, we see that the additional Kerr nonlinearity $\beta=0.2$ does not change the synchronization boundary.}
\end{figure}

As a measure for synchronization enhancement, we can calculate the `total bandwidth' by integrating the synchronization bandwidth from $\eta = 0$ to some $\eta = \eta_\text{max}$, which can then be evaluated asymptotically for large $\eta_\text{max}$ (compared to $\lambda$). This gives (assuming $\eta_\text{max} > \lambda$)
\begin{multline}
    B(\lambda,\eta_\text{max}) = \int_0^\lambda d\eta \, \eta + \int_\lambda^{\eta_\text{max}} d\eta \, \sqrt{\lambda (2\eta - \lambda)} \\
    = \frac{1}{6}\lambda^2 + \frac{1}{3} \, \lambda^{1/2} \, (2 \eta_\text{max} - \lambda)^{3/2} \approx \frac{2\sqrt{2}}{3} \, \lambda^{1/2} \, \eta_\text{max}^{3/2}  \;,
\end{multline}
which increases as $\lambda^{1/2}$. Since $B$ has the geometrical interpretation of the area covered by the synchronization region in the parameter space $(\lambda,\eta)$, this means that the synchronization region grows with the vdP nonlinearity $\lambda$.

\section{Two classical oscillators with reactive coupling}

Let us first consider two classical SL oscillators which are reactively coupled with strength $g$. The coupled complex-amplitude equations are
\begin{equation}
\begin{split}
    \alpha'_1 &= -i \, \alpha_1 + \frac{\lambda}{2} (1-|\alpha_1|^2) \,\alpha_1 + i \, g (\alpha_2 - \alpha_1)  \;,  \\ 
    \alpha'_2 &= -i\,(1+\Delta) \,\alpha_2 + \frac{\lambda}{2} (1-|\alpha_2|^2) \,\alpha_2 + i \, g (\alpha_1 - \alpha_2) \;.
\end{split}    
\end{equation}
Similar to before, we work in polar coordinates, giving
\begin{equation}
\begin{split}
    R'_1 &= \frac{\lambda}{2} \, R_1 (1-R_1^2) + g \, R_2 \sin\varphi  \;,\\ 
    R'_2 &= \frac{\lambda}{2} \, R_2 (1-R_2^2) - g \, R_1 \sin\varphi  \;, \\ 
    \varphi' &= \Delta + g \left( \frac{R_2}{R_1} - \frac{R_1}{R_2} \right) \cos\varphi \;.
\end{split}
\end{equation}
At steady state, $R_1 = R_2$, and we see that $\varphi' = \Delta$. In other words, the rate at which the phase difference grows is exactly the detuning between the two oscillators. Thus for reactive coupling, the two oscillators simply cannot synchronize. This result generalizes to the case two quantum SL oscillators.

\begin{figure}
\centering
\subfloat{%
  \includegraphics[width=\linewidth]{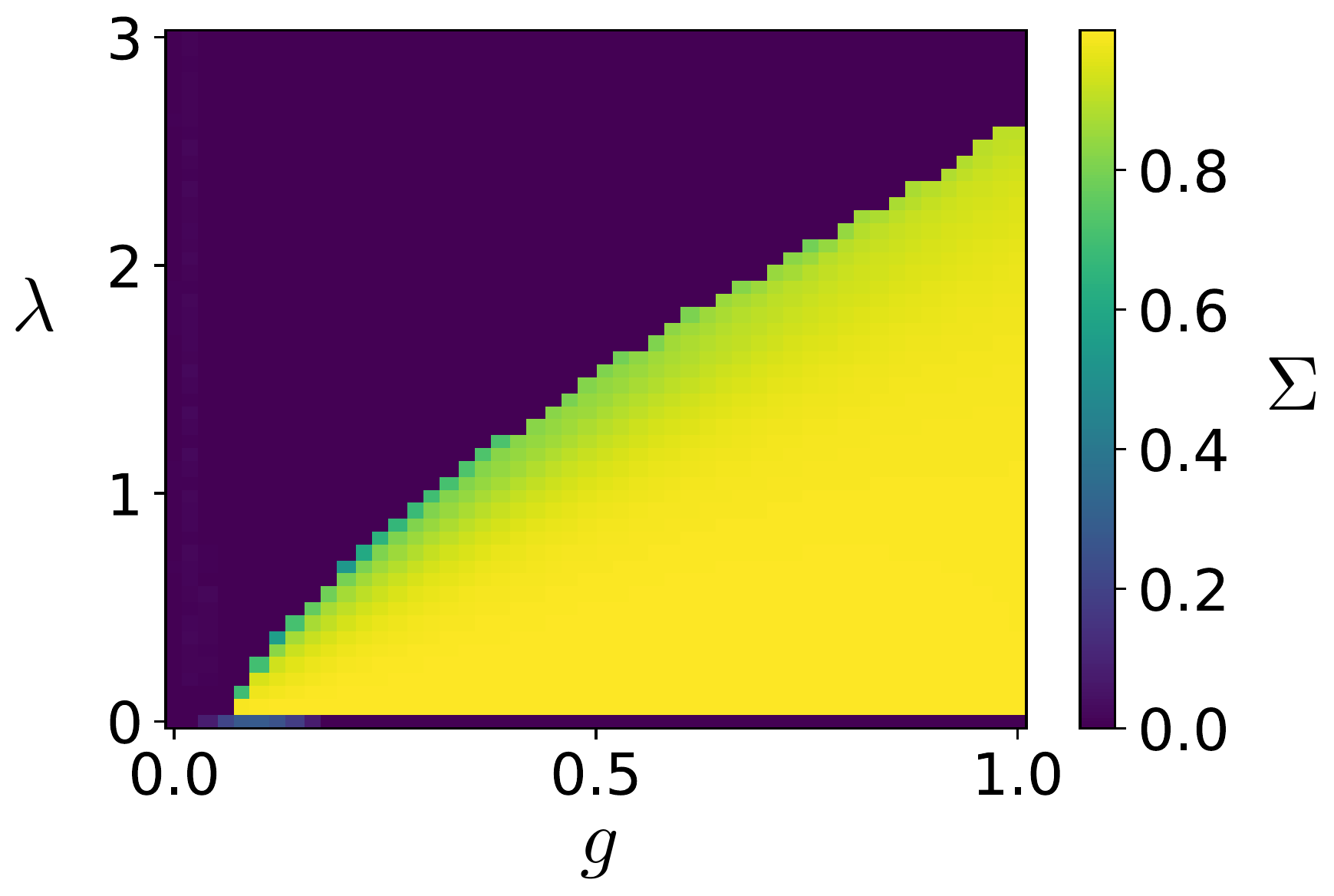}%
  \label{}%
}
	\caption{Correlation coefficient of two reactively-coupled vdP oscillators for $\Delta = 0.1, r = 1$.}
	\label{fig:2vdP_reactive}
	\end{figure}

What is particularly interesting in the case of reactive coupling is how the oscillators behave beyond the $\lambda\longrightarrow0^+$ limit. In the main text we have shown that increasing the nonlinearity in two reactively coupled quantum vdP oscillators increases their position correlation before a plateau is reached. This result is particularly interesting because the analogous classical system fails to produce the same effect. Here we show this explicitly by computing the steady-state correlation coefficient for the positions of two classical vdP oscillators as a function of their nonlinearities (assumed identical, given by $\lambda$) and their coupling strength. The position correlation coefficient in this case is given by
\begin{align}
\label{sigma(x1,x2)}
    \Sigma = \frac{\sum_{m=1}^M \big[ x_1^{(m)}-\bar{x}_1 \big] \big[ x_2^{(m)}-\bar{x}_2 \big]}{\sqrt{\sum_{m=1}^M \big[ x_1^{(m)}-\bar{x}_1 \big]^2} \sqrt{\sum_{m=1}^M \big[ x_2^{(m)}-\bar{x}_2 \big]^2}}  \;,
\end{align} 
where $\big\{\big(x_1^{(1)},x_2^{(1)}\big),\big(x_1^{(2)},x_2^{(2)}\big),\ldots,\big(x_1^{(M)},x_2^{(M)}\big) \big\}$ are pairwise samples of $\big(x_1(t),x_2(t)\big)$. Note that since we are interested in how $x_1(t)$ and $x_2(t)$ are correlated in the long-time time limit, each pairwise sample must be taken from $x_1(t)$ and $x_2(t)$ after all transience has died out. We have also defined
\begin{align}
    \bar{x}_k = \frac{1}{M} \, \sum_{m=1}^M  x_k^{(m)}  \;.
\end{align}
For the two classical vdP oscillators we may simply take  $\big(x_1^{(m)},x_2^{(m)})\big)=\big(x_1(m\,\delta t),x_2(m\,\delta t)\big)$ where we have introduced a small time increment $\delta t$. For a sufficiently large sample size $M$, \eqref{sigma(x1,x2)} measures how correlated $x_1(t)$ and $x_2(t)$ are in the long-time limit (or steady state). The result of such a computation is shown in Fig.~\ref{fig:2vdP_reactive} as a function of $\lambda$ and $g$.
As $\lambda \longrightarrow 0$, we can see that the correlation vanishes, as expected from the SL model. But more importantly, and omitting the degenerate $\lambda = 0$ case, we find that $\Sigma$ appears to decrease monotonically with $\lambda$, and decreases sharply to zero when $\lambda$ exceeds some critical value. It is simply impossible to increase $\Sigma$ by increasing $\lambda$ for two reactively-coupled classical vdP oscillators at a fixed $g$. This is in stark contrast to the same calculation performed for two such quantum vdP oscillators for which $\Sigma$ can increase if $\lambda$ is increased for a given $g$.


\section{Quantum synchronization in the deep quantum limit}

\subsection{Dissipatively-coupled oscillators}

We derive here a sufficient condition for amplitude death in two dissipatively coupled SL oscillators in the deep quantum regime. The density operator for two dissipatively-coupled oscillators satisfies $\rho'=\Lcal\rho$ where $\Lcal$ given by
\begin{multline}
    \Lcal = -i\,\Delta \, \big[ \hat{a}_1^\dag \hat{a}_1,\supopdot\, \big] + \kappa \, \big( \Dcal[\hat{a}_1^\dag] + \Dcal[\hat{a}_2^\dag] \, \big) + \gamma \, \big( \mathcal{D}[\hat{a}_1^2] + \Dcal[\hat{a}_2^2] \, \big) \\
    + \eta \,\Dcal[\hat{a}_1-\hat{a}_2]  \;,
\end{multline}
where $\Delta$ is the detuning between the two oscillators and $\eta$ is the dissipative coupling strength. Note that we have assumed equal single-photon amplification rates $\kappa$, and equal two-photon dissipation rates $\gamma$ for the two oscillators. In the deep quantum limit, i.e.~when $\gamma/\kappa \longrightarrow \infty$, the process of two-photon loss dominates and this confines each oscillator to the zero- and one-photon subspace. This allows us to treat each oscillator effectively as a two-level system. After adiabatically eliminating the higher excited states \cite{lee2014/physreve.89.022913}) we have
\begin{multline}
\bar{\Lcal }= -i \, \bar{\Delta} \,[\hat{\sigma}_1^+ \hat{\sigma}_1^-,\supopdot \,] + \Dcal[\hat{\sigma}_1^+] + \Dcal[\hat{\sigma}_2^+] + 2 \, \Dcal[\hat{\sigma}_1^-] + 2 \, \Dcal[\hat{\sigma}_2^-] \\
+ \bar{\eta} \, \Dcal[\hat{\sigma}_1-\hat{\sigma}_2^-]
\end{multline}
where $\bar{\Delta} \equiv \Delta/\kappa$ and $\bar{\eta} \equiv \eta/\kappa$. We have also denoted the creation and annihilation operators for the $j$th oscillator in the $\{|0\rangle,|1\rangle\}$ subspace by $\hat{\sigma}_j^+$ and $\hat{\sigma}_j^-$. With this simplification, we can solve for the steady-state density matrix exactly, defined by $\bar{\Lcal}\varrho=0$. In the basis $\{\ket{00},\ket{01},\ket{10},\ket{11}\}$ we find
\begin{equation}
\varrho = \begin{pmatrix} \varrho_{11} & 0 & 0 & 0 \\ 0 & \varrho_{22} & \varrho_{23} & 0 \\ 0 & \varrho_{32} & \varrho_{33} & 0 \\ 0 & 0 & 0 & \varrho_{44} \end{pmatrix}  \;.
\end{equation}
The matrix elements are given explicitly by
\begin{equation}
\begin{split}
    \varrho_{11} &= \big[ 6 \, \bar{\eta}^3 + (\bar{\eta}+2)^2 \bar{\Delta}^2 + 34 \, \bar{\eta}^2 + 60 \, \bar{\eta} + 36 \big]/\nu  \;,  \\
    \varrho_{22} = \varrho_{33} &= \big[ \bar{\eta}^3 + (\bar{\eta}+2)^2 \, \bar{\Delta}^2 + 8 \bar{\eta}^2 + 21 \bar{\eta} + 18 \big]/\nu   \;,  \\
    \varrho_{44} &= \big[ \bar{\eta}^2 + \bar{\Delta}^2 + 6 \, \bar{\eta} + 9 \big]/\nu  \;, \\
    \varrho_{23} = \varrho_{32}^* &= \big[ \bar{\eta} (\bar{\eta}+1)(\bar{\eta}+3) + i \, \bar{\eta} (\bar{\eta}+1) \bar{\Delta} \big]/\nu    \;,
\end{split}
\end{equation}
where for ease of writing we have introduced the factor 
\begin{align}
    \nu = 8 \, \bar{\eta}^3 + (\bar{\eta}+3)^2 \bar{\Delta}^2 + 51 \, \bar{\eta}^2 + 108 \, \bar{\eta} + 81 \;.
\end{align}
It is straightforward to evaluate the position correlation coefficient     
\begin{align}
     \Sigma = {}& \frac{\braket{\hat{x}_1 \hat{x}_2}-\braket{\hat{x}_1}\braket{\hat{x}_2}}{\sqrt{\big[ \braket{\hat{x}_1^2}-\braket{\hat{x}_1}^2 \big] \big[\braket{\hat{x}_2^2}-\braket{\hat{x}_2}^2\big]}}  \nn \\
    = & \frac{\rho_{23}+\rho_{32}}{{\rm Tr}[\varrho]} = \frac{2\,\bar{\eta}\,(\bar{\eta}+1)}{8\,\bar{\eta}^2 + 27\,\bar{\eta}  + (\bar{\eta}+3)\bar{\Delta}^2 + 27}
\end{align}
where $\hat{x}_j = \hat{\sigma}_j^+ + \hat{\sigma}_j^-$ ($j=1,2$). The maximum correlation $\Sigma \longrightarrow 1/4$ is achieved in the limit $\bar{\Delta} \longrightarrow 0$ and $\bar{\eta} \longrightarrow \infty$. To study amplitude death, we trace out the second oscillator, giving the single-oscillator density matrix
\begin{equation}
    \varrho_1 = {\rm Tr}_2[\varrho] = \begin{pmatrix} \varrho_{11} + \varrho_{22} & 0 \\ 0 & \varrho_{33} + \varrho_{44} \end{pmatrix}.
\end{equation}
Its Wigner distribution, taken as a function of the radial coordinate $r$, can then be calculated explicitly as  
\begin{equation}
    W_1(r) = (\varrho_{11} + \varrho_{22}) \, e^{-r^2} + (\varrho_{33}+\varrho_{44}) (2\,r^2-1) \, e^{-r^2}  \;.
\end{equation}
We define amplitude death to be the regime where the Wigner distribution possesses a single peak at the origin instead of being ring like. A sufficient and necessary condition for this is when $\varrho_{11}+\varrho_{22} \geq 3/4$, or equivalently,
\begin{equation}
    \bar{\Delta}^2 \geq \frac{27-15\,\bar{\eta}^2 - 4\,\bar{\eta}^3}{(\bar{\eta}-1)(\bar{\eta}-2)}  \;. 
\end{equation}
Interestingly, when the two oscillators are on resonance ($\bar{\Delta}=\Delta=0$), amplitude death can still occur as long as $\bar{\eta} \gtrsim 1.17$. This is due to the quantum noise which is significant in the deep quantum limit, rather than coming from the frequency mismatch between the oscillators.

\subsection{Reactively-coupled oscillators}

We show here that two identical reactively-coupled quantum SL oscillators cannot share any position correlations. This permits calling the correlations observed in two similarly coupled vdP oscillators to be nonlinearity induced. For two reactively-coupled quantum SL oscillators, its Lindbladian (in units where $\kappa=1$, $\Delta=0$)
\begin{equation}
\Lcal = -ig \, \big[\hat{a}_1^\dag \hat{a}_2 + \hat{a}_2^\dag \hat{a}_1,\supopdot\,\big] + \mathcal{D}\big[\hat{a}_1^\dag] + \mathcal{D}\big[\hat{a}_2^\dag\big]  + \gamma \, \big( \mathcal{D}\big[\hat{a}_1^2] + \mathcal{D}\big[\hat{a}_2^2\big] \big)  \;.
\end{equation}
To order $1/\gamma$, we can truncate the Hilbert space of each oscillator to the first three levels, and solve for the steady-state density matrix $\varrho$ (again defined by $\Lcal\varrho=0$), 
\begin{equation}
\begin{split}
    \varrho =& \left[\frac{4}{9}+\frac{4(7g^2-6)}{81\gamma}\right] \ket{00}\bra{00} \\
    &+ \left[\frac{2}{9} - \frac{4(g^2+3)}{81\gamma}\right] (\ket{01}\bra{01}+\ket{10}\bra{10}) \\
    &+ \frac{2}{9\gamma} \ket{02}\bra{02} + \frac{2}{9\gamma} (\ket{02}\bra{02} + \bra{20}\bra{20}) \\& + \left[\frac{1}{9} - \frac{2(10g^2+3)}{81\gamma}\right] \ket{11}\bra{11} \\
    &+ \frac{1}{9\gamma} \big( \ket{12}\bra{12}+\ket{21}\bra{21} \big) \\
    &+ \frac{i\sqrt{2}g}{9\gamma} \big( \ket{11}\bra{20} + \ket{11}\bra{02} - \ket{02}\bra{11} - \ket{20}\bra{11} \big) \;.
\end{split}
\end{equation}
It can then be checked explicitly from this expression for $\varrho$ that $\Sigma$ is always zero.

\bibliography{ref}

\end{document}